\documentclass[12pt]{article}

\usepackage{amssymb,amsmath,amsfonts,eurosym,geometry,ulem,graphicx,caption,color,setspace,sectsty,comment,footmisc,caption,pdflscape,subfigure,array}
\usepackage[short, nocomma]{optidef}
\usepackage{multirow}
\usepackage[ruled,vlined]{algorithm2e}
\usepackage{enumitem}
\usepackage{longtable}
\usepackage{adjustbox}
\usepackage[mathscr]{eucal} 
\usepackage[short, nocomma]{optidef}

\usepackage[section]{placeins} 
\usepackage[hidelinks]{hyperref}

\usepackage{booktabs}
\usepackage{indentfirst}
\usepackage{float}

\usepackage[
  style= ieee,
  natbib=true,%
  clearlang=true,%
  backend=biber,%
]{biblatex}

\AtEveryBibitem{%
  \clearfield{day}%
  \clearfield{month}%
  \clearfield{endday}%
  \clearfield{endmonth}%
  \clearfield{note}
  \clearfield{urlyear}
  \clearfield{location}
  \clearfield{publisher}
}

\DeclareRedundantLanguages{en,EN,English}{english}

\DeclareFieldFormat{pages}{\mkfirstpage{#1}}

\newcolumntype{L}[1]{>{\raggedright\let\newline\\arraybackslash\hspace{0pt}}m{#1}}
\newcolumntype{C}[1]{>{\centering\let\newline\\arraybackslash\hspace{0pt}}m{#1}}
\newcolumntype{R}[1]{>{\raggedleft\let\newline\\arraybackslash\hspace{0pt}}m{#1}}

\begin{document}

\begin{titlepage}
\title{Auction designs to increase incentive compatibility and reduce self-scheduling in electricity markets}
\author{Conleigh Byers\thanks{C. Byers is with Harvard University in Cambridge, MA, USA (email: \url{cbyers@fas.harvard.edu}), formerly ETH Z{\"u}rich in Z{\"u}rich, Switzerland and is the corresponding author} \and Brent Eldridge\thanks{B. Eldridge is with Pacific Northwest National Laboratory (PNNL) in Richland, WA, USA (email: \url{brent.eldridge@pnnl.gov})}}
\date{March 15, 2023}
\maketitle
\begin{abstract}
\noindent The system operator's scheduling problem in electricity markets, called unit commitment, is a non-convex mixed-integer program. The optimal value function is non-convex, preventing the application of traditional marginal pricing theory to find prices that clear the market and incentivize market participants to follow the dispatch schedule. Units that perceive the opportunity to make a profit may be incentivized to self-commit (submitting an offer with zero fixed operating costs) or self-schedule their production (submitting an offer with zero total cost). We simulate bidder behavior to show that market power can be exercised by self-committing/scheduling. Agents can learn to increase their profits via a reinforcement learning algorithm without explicit knowledge of the costs or strategies of other agents. We investigate different non-convex pricing models over a multi-period commitment window simulating the day-ahead market and show that convex hull pricing can reduce producer incentives to deviate from the central dispatch decision. In a realistic test system with approximately 1000 generators, we find strategic bidding under the restricted convex model can increase total producer profits by 4.4\% and decrease lost opportunity costs by 2/3. While the cost to consumers with convex hull pricing is higher at the competitive solution, the cost to consumers is higher with the restricted convex model after strategic bidding. \\
\vspace{0in}

\bigskip
\end{abstract}
\setcounter{page}{0}
\thispagestyle{empty}
\end{titlepage}
\pagebreak \newpage

\doublespacing

\section{Introduction}

Price formation efforts in wholesale electricity markets are premised on assumptions of competitive market behavior such that all participants are price takers in the spot market. In short, the consequence is that offers submitted to the ISO reflect the actual marginal cost of each resource and the market clearing price is set by the marginal cost of the highest cost offer that the ISO accepts. The impact of these assumptions can be widespread: for example, studies on long-term investment often take this aspect of the spot market for granted. It is therefore important to critically assess whether wholesale electricity price formation policies currently support competitive behavior in the spot markets.

A growing literature on non-convex pricing has highlighted the absence of uniform market-clearing prices in practical wholesale electricity market scheduling problems \cite{liberopoulos_critical_2016,herrero_electricity_2015,mays_investment_2020,byers_long-run_2022}. In addition to marginal production costs, conventional thermal generators also have avoidable fixed costs relating to their start-up, shut-down, and operating status, and opportunity costs related to minimum production level when they are online and the minimum up-time or down-time between start-up and shut-down decisions. The issue of non-convexities arises in markets that solve a mixed integer liner program (MILP) called the security constrained unit commitment (SCUC) problem to efficiently schedule conventional thermal generators during the day-ahead market \cite{hobbs2006next}, which is commonly implemented in the United States. An alternate market design in which participants attempt to internalize their non-convex costs in block orders (leading yet still to a non-convex problem for the market operator) is common in Europe \cite{van_vyve_linear_2011}.  It is typically not possible to determine a uniform market clearing price where all market participants are able to maximize their profit by following the socially optimal production schedule determined by the system operator. 

The system operator solves an optimization problem with the operating constraints of the units and calculates uniform prices that are charged to all participants in the auction. Payments typically also include side payments to individual units to ensure that they suffer no short-run losses from following the central dispatch decision. However, ensuring no short-run losses does not guarantee that units will have no lost opportunity costs. Lost opportunity costs are the difference between a generator's preferred profit achieved when producing to maximize its profit in response to the price and the profit achieved when following the system operator's socially-optimal dispatch schedule. While the exercise of market power by offering untruthful bids is a concern in many markets \cite{graf_market_2021}, we show that that these lost opportunity costs may motivate market participants to bid strategically to improve their outcomes by offering their desired production quantity at zero cost. A generator in a non-convex market is able to exercise market power by self-committing/scheduling.

An ideal pricing mechanism achieves four properties. First, it achieves market efficiency by maximizing social welfare and resulting in an outcome from which no participant wishes to unilaterally deviate. Second, participants should recover their variable costs (although not necessarily their fixed capital costs) in the short-run. Third, it is revenue adequate. The amount of revenue recovered from consumers is at least as great as
the amount of revenue paid to suppliers. Finally, the ideal pricing mechanism is incentive compatible: participants do best when offering their true preferences or costs. Each producer maximizes its own payoff by bidding its true supply costs, and no participants have an incentive to exercise market power by bidding strategically.

If the market is convex, i.e., the optimal value function seen by the market operator is convex, and participants must bid their true costs, then pricing at marginal cost yields an outcome that achieves market efficiency, cost recovery, and revenue adequacy. No participant faces a lost opportunity cost. The optimality conditions for the equilibrium market problem in which each participant seeks to optimize its individual benefit and the system operator's optimization problem seeking to optimize social welfare are equivalent, and thus the social-welfare maximizing outcome is the same as the market outcome. However, if the market is not perfectly competitive, it is possible for participants to bid strategically and increase their payoffs. Thus, marginal pricing does not guarantee incentive compatibility if the operator has imperfect information and producers can increase their supply offers above their true costs. In fact, no market-clearing mechanism ensures all four properties at the same time \cite{hurwicz1972informationally, myerson1983efficient}. A trade-off must be made, and alternative pricing methods may achieve different properties. While it is possible to ensure incentive compatibility with the the Vickrey–Clarke–Groves (VCG) mechanism, in which truthful bidding is the dominant strategy \cite{xu_efficient_2017}, revenue adequacy is no longer guaranteed, although strategies have been proposed to reduce the market operator's budget deficit \cite{xu_efficient_2017, hobbs_evaluation_2000}.

If a market has non-convex costs, often no uniform price can be found that supports the market operator's schedule, resulting in significant lost opportunity costs. A number of methods for pricing in the presence of non-convexities have been proposed. Authors in \cite{oneill_efficient_2005} propose relaxing integrality and fixing binary variables to the previously-found optimal values, a method we will call fixed configuration pricing (FCP). This can result in instances in which the generator that sets the price does not have the highest variable costs, and thus the price may decrease as demand increases. Lost opportunity costs may be high, i.e., generators may not be incentivized to follow the central dispatch decision. Another proposal by \cite{gribik_market-clearing_2007} called convex hull pricing (CHP) seeks to find a uniform price that minimizes lost opportunity costs. There is evidence this approach improves long-run incentives \cite{herrero_electricity_2015,azizan_optimal_2020,mays_investment_2020,byers_long-run_2022}, and \cite{andrianesis_computation_2021} propose a new computationally tractable method using Dantzig-Wolfe decomposition to find exact convex hull prices. 

While attention in incentive compatibility discussions is primarily given to economic offers \cite{baldick2004theory,sioshansi2007good,hortaccsu2008understanding}, non-convex markets raise the possibility of increasing payoffs by submitting zero-cost supply offers for the desired level of production. A stylized test case proposed in \cite{eldridge_pricing_2018,eldridge_algorithms_2020}, replicated in Section \ref{sec:stylizedtestcase}, demonstrates that market power in non-convex markets can be exercised by self-committing. Some Nash equilibria strategies include zero-cost supply offers.

In electricity markets, the phenomenon of self-committing or self-scheduling by submitting offers below actual costs is widespread. A self-commitment is when a generator indicates to the system operator that it wishes to be dispatched at least at its minimum operating level regardless of the market price. From the operator's perspective, this equates to submitting an offer with zero costs up to the minimum operating level. Similarly, a self-schedule entails submitting an offer with zero costs up to the desired dispatch quantity. Self-commitment and self-schedules constitute approximately 40\% of the energy market offers in the PJM market \cite{monitoring_analytics_2018_2019}. We can characterize a self-commitment or self-schedule offers as benign or adverse, depending on whether it reduces market efficiency. Benign offers may be submitted if a resource's startup or notification time exceeds the window of the 24-hour day-ahead market, to avoid transaction costs of gathering cost information for units that are very likely to be dispatched, or because of take-or-pay fuel contracts that render some portion of the generator's output a sunk cost. However, an adverse offer would result in greater profits in expectation for a generator than if the generator had submitted an economic bid reflective of its true costs.  It is unclear how many (if any) adverse self-commitments and self-schedules exist, as such a strategy may be difficult to detect by conventional market power mitigation software. Nevertheless, adverse self-commitment and self-schedules could result in lower market efficiency if they cause the system operator to find a suboptimal dispatch decision due to the distorted costs. To disincentivize inefficient behavior, generators that self-commit or self-schedule are typically not eligible for make-whole payments. Notably, the type of units found to strategically self-commit in \cite{eldridge_algorithms_2020} share similar characteristics to the coal generators that often self-commit in reality \cite{miso_miso_2020}, namely that they are "inflexible, relatively expensive, and mostly profitable" \cite{eldridge_algorithms_2020}.

Authors in \cite{eldridge_pricing_2018,eldridge_algorithms_2020} demonstrate the ability of generators to strategically self-commit in a stylized test system with a single operating time period. The question remains as to whether generators could determine optimal strategies in a more realistic system with many different generator attributes across a multi-period optimization horizon. Self-commitment and self-scheduling allows ``out-of-merit” resources to enter the SCUC solution yet remain profitable; each out-of-merit commitment and dispatch may cause a cascading change in market prices and the commitment and dispatch of other resources in the market. The outcome cannot be be explicitly modeled by individual participants in realistically sized markets. To avoid this issue, we show that market agents can implicitly identify profitable self-commitment and -scheduling strategies via simple reinforcement learning algorithms, i.e., without using a sophisticated model for how an agent’s self-commitment or self-scheduling will affect the SCUC solution. We investigate the ability of participants to learn to improve outcomes by self-scheduling or self-committing via a reinforcement learning algorithm in a large-scale test system over an operating day, simulating a day-ahead market. We examine two competing pricing models, showing that the ability of generators to adversely self-commit or self-schedule is decreased with convex hull pricing.

\section{Pricing Models}\label{sec:sspPM}

A simple unit commitment problem with variable cost $C$, startup cost $F$, production $p$, and commitment status $u$ linked to startup decision $z$ with inelastic demand is formulated as: 

\begin{mini!}[2]<>
{u,p \in \mathscr{P}}{\sum_{t\in T}\sum_{g\in G}{(C_g p_{tg} + F_g z_{tg})}  } {}{}
\addConstraint{\sum_{g\in G}{p_{tg}}}{= D }{\quad \forall t \in T  }
\addConstraint{u}{\in \{0,1\} \label{eq:SS:UCbin}}
\end{mini!}

where $\mathscr{P}$ is the set of operating constraints.

For a price signal ${\lambda^*}$ to incentivize an agent to follow the system operator's dispatch decision ${p^*}$, the production quantity must solve a function that maximizes each generator’s profits given ${\lambda^*}$:
\begin{maxi}[2]<>
{p_g,u_g}{\Pi_{g}(\lambda^{*},p_g,u_g) \quad \forall g\in G} {}{}
\end{maxi}

With market-clearing prices $\lambda$, variable costs $C$, startup costs $F$, startup decision $z$ linked to commitment status $u$, and side payments $s$, initial short-run profit $\Pi^0$ and final short-run profit $\Pi$ are defined as:

\begin{equation}
\Pi^0_g := \sum_{t \in T}((\lambda_t - C_g)p_{tg} - F_g z_{tg})
\end{equation}
\begin{equation}
\Pi_g := \Pi_g + \sum_{t \in T}s_{tg}
\end{equation}

Perceived losses are lost opportunity costs (LOC), the difference between what a unit could make given a price if able to schedule its own dispatch (its preferred profit) and what it would make with the same price following the centralized dispatch decision, plus any additional compensation received as side payments. We define initial lost opportunity costs $LOC^0$ without side payments and final $LOC$ as:
\begin{maxi}[2]<>
{u_g,p_g}{\Pi_{g}^0(\lambda^{*},u_g,p_g) - \Pi_{g}^0(\lambda^{*},u_g^*,p_g^*)} {}{LOC_g^0 :=}
\end{maxi}
\begin{equation}
LOC_g := LOC_g^0 + \sum_{t \in T}s_{tg}
\end{equation}

Provided the unit could choose to not produce, a subset of $LOC^0$ is make-whole payments (MWP), the revenue required for short-run cost recovery. This occurs when the unit would prefer to not operate at the market-clearing price. MWP are typically determined for the same timescale at the day-ahead market, i.e., for each 24-hour period:
\begin{equation}
MWP_g := - \min(0,\Pi_g^0(\lambda^*,u^*,p^*))
\end{equation} 

Prices cannot be derived directly from the unit commitment problem without relaxation. The method proposed in \cite{oneill_efficient_2005} fixes the binary variables in the to their optimal values and then computes prices $\lambda$ from the Lagrangian multipliers of the resulting linear program. We call this method fixed configuration pricing (FCP). 

\begin{mini!}[2]<>
{u,p \in \mathscr{P}}{\sum_{t\in T}\sum_{g\in G}{(C_g p_{tg} + F_g z_{tg})}  } {}{}
\addConstraint{\sum_{g\in G}{p_{tg}}}{= D }{\quad \forall t \in T \quad : \lambda_t  }
\addConstraint{u = u^*}{}{}
\addConstraint{u}{\in [0,1] \label{eq:SS:UCbinSS}}
\end{mini!}

An alternative model seeks to find the uniform price that minimizes lost opportunity costs, deriving prices from the convex hull of the optimal value function. In convex hull pricing (CHP), prices are determined by solving the Lagrangian dual of the UC problem. Approximate CHP (aCHP) can be calculated by identifying a close approximation of the convex hull of the primal UC problem \cite{hua_convex_2017}. If a good approximation can be found, aCHP prices $\lambda$ are given by simply relaxing the binary variables: 

\begin{mini!}[2]<>
{\{u,p\} \in \mathscr{P}}{\sum_{t\in T}\sum_{g\in G}{(C_g p_{tg} + F_g z_{tg})}  } {}{}
\addConstraint{\sum_{g\in G}{p_{tg}}}{= D }{\quad \forall t \in T \quad : \lambda_t }
\addConstraint{u}{\in [0,1] }
\end{mini!}

\section{Methodology}

\subsection{Unit Commitment Model}\label{sec:UC_model}

The unit commitment problem to be solved in each iteration is given below. The short-run profit for each generator $g$ is:

\begin{equation}
\Pi^0_g := \sum_{t \in T}(\lambda_t p_{tg} - \sum_{s \in S}{C_{gs} \rho_{tgs}} -H_g u_{tg} - F_g z_{tg})
\end{equation}

\subsubsection{Nomenclature} 

\begingroup
\begin{longtable}[l]{ll}
    \multicolumn{2}{l}{\textit{Indices and Sets}} \\[5pt]
    $g \in G$ &Set of generators  \\
    $G^T \subseteq G$ &Set of thermal generators  \\
    $G^V \subseteq G$ &Set of VRE resources  \\
    $t \in T$ &Set of time periods (hours) \\
    $s \in S$ &Set of offer steps
    \\
    \multicolumn{2}{l}{\textit{Parameters}} \\[5pt]
    $C_{gs}$ &Variable cost in offer step $s$ (\$/MWh)\\
    $F_g$ &Startup cost (\$)\\
    $H_g$ &No load cost (\$) \\
    $P_g^{min}$ &Minimum operating capacity (MW)\\
    $P_{gs}^{max}$ &Maximum operating capacity of offer step $s$ (MW)\\
    $M_g^{on}$ &Minimum on time (h) \\
    $M_g^{off}$ &Minimum off time (h)\\
    $R_g^{+}$ &Maximum ramp up rate (MW/h)\\
    $R_g^{-}$ &Maximum ramp down rate (MW/h)\\
    $\mathcal{P}_{tg}$ &Maximum output for VRE resource (MW) \\
    $D_{t}$ &Maximum quantity of demand bids at time $t$ (MW)\\
    $U_{g}^{init}$ & Initial status of generator (Binary)\\
    \\
    \multicolumn{2}{l}{\textit{Variables}} \\[5pt]
    $p_{tg}$ & Committed generation for generator $g$ at time $t$ (MW)\\
    $\rho_{tgs}$ & Generation for generator $g$ in offer step $s$ at time $t$ (MW)\\
    $u_{tg}$ & (Binary) commitment status for generator $g$ at time $t$\\
    $z_{tg}$ & (Binary) startup decision for generator $g$ at time $t$\\
    $y_{tg}$ & (Binary) shutdown decision for generator $g$ at time $t$\\
    $n_{t}$ &Non-served energy at time $t$ (MW)\\
\end{longtable} 
\endgroup
\addtocounter{table}{-1}

\subsubsection{Formulation}

\begin{mini!}[3]<b>
{(p,\rho,u,z,y)}{ \sum_{t\in T}\sum_{g\in G}{( \sum_{s \in S}(C_{gs} \rho_{tgs}) + H_g u_{tg} + F_g z_{tg})} + \sum_{t\in T}{n_t} } {\label{eq:SS:UC}}{}
\addConstraint{\sum_{g\in G}{p_{tg}} + n_t = D_{t} }{}{ \forall t \in T}\label{eq:SS:PowerBalancePRD}
\addConstraint{\sum_{s \in S}\rho_{tgs} = p_{tg}}{}{\forall t \in T, g \in G}
\addConstraint{z_{tg} + y_{tg} \leq 1 }{}{\forall t \in T, g \in G^T} \label{eq:SS:OnOff1}
\addConstraint{u_{tg} - u_{t-1,g} = z_{tg} - y_{tg} }{}{\forall t \in 2,...,T, g \in G^T} \label{eq:SS:OnOff2}
\addConstraint{z_{tg} = u_{tg} }{}{\forall t = 1, g \in G^T: U_{g}^{init}=0} \label{eq:SS:StartStateoff}
\addConstraint{z_{tg} = 0 }{}{\forall t = 1, g \in G^T: U_{g}^{init}=1} \label{eq:SS:StartStateon}
\addConstraint{z_{tg}+\sum_{t'=t+1}^{\min{(t+M_g^{on}-1,T)}}{y_{t'g}} \leq 1 }{}{} \label{eq:SS:Minonoff1} \nonumber
\addConstraint{}{}{\forall t \in 1,...,T-1, g \in G^T: M_g^{on} > 1 }
\addConstraint{y_{tg} + \sum_{t'=t+1}^{\min{(t+M_g^{off}-1,T)}}{z_{t'g}} \leq 1 }{}{}\label{eq:SS:Minonoff2} \nonumber
\addConstraint{}{}{\forall t \in 1,...,T-1, g \in G^T: M_g^{off} > 1}
\addConstraint{P_g^{min} u_{tg} \leq p_{tg}}{}{\forall t \in T, g \in G^T} \label{eq:SS:CapCon}
\addConstraint{\rho_{tgs} \leq P_{gs}^{max} u_{tg}  }{}{\forall t \in T, g \in G^T, s \in S}
\addConstraint{0 \leq p_{tg} \leq \mathcal{P}_{tg} }{}{\forall t \in T, g \in G^V}\label{eq:SS:VRE}
\addConstraint{p_{tg} \geq 0  }{}{\forall t \in T, g \in G}
\addConstraint{u_{tg},z_{tg},y_{tg} \in \{0,1\}  }{}{\forall t \in T, g \in G^T}
\addConstraint{x_g \in \{0,1\} }{}{\forall g \in G^T} \label{eq:SS:Build}
\end{mini!}

We assume generators can ramp up to or down from $P^{min}$ during startup or shutdown. We use tight UC constraints in attempt to better approximate the convex hull of the optimal value function when relaxing binary variables \cite{knueven2018ramping}. Ramping constraints are implemented as the two-period ramp inequalities proposed in \cite{damci2016polyhedral}:

\begin{align}
    p_{tg} \leq p_{t-1,g} + (P_g^{min} + R_g^{+})u_{tg} - P_g^{min} u_{t-1,g} - R_g^{+} z_{tg} \nonumber \\
    \forall t = 2,...,T, g \in G^T \label{eq:SS:TightRamp1} \\
    p_{t-1,g} \leq p_{t,g} + (P_g^{min} + R_g^{-})u_{t-1,g} - P_g^{min} u_{t,g} - R_g^{-} y_{tg} \nonumber \\
    \forall t = 2,...,T, g \in G^T \label{eq:SS:TightRamp2} 
\end{align}

Let $P_g^{MAX} = \sum_{s \in S}P_{gs}^{max}$, the maximum capacity of generator $g$. We implement the following constraints from \cite{gentile2017tight} that serve only to tighten the UC formulation:

\begin{align}
    p_{tg} \leq  P_g^{MAX} u_{tg} - (P_g^{MAX} - P_g^{min}) y_{t+1,g}  \nonumber\\
    \forall t=1, g \in G^T\\
    p_{tg} \leq P_g^{MAX} u_{tg} - (P_g^{MAX} - P_g^{min}) z_{t,g} - (P_g^{MAX} - P_g^{min}) y_{t+1,g} \nonumber\\
    \forall t=2,...,T-1, g \in G^T : M_g^{on} \geq 2 \\
    p_{tg} \leq P_g^{MAX} u_{tg} - (P_g^{MAX} - P_g^{min}) z_{t,g} \nonumber\\
    \forall t=T, g \in G^T\\
    p_{tg} \leq P_g^{MAX} u_{tg} - (P_g^{MAX} - P_g^{min}) z_{t,g} \nonumber\\
    \forall t=2,...,T-1, g \in G^T : M_g^{on} < 2 \\
    p_{tg} \leq P_g^{MAX} u_{tg} - (P_g^{MAX} - P_g^{min}) y_{t+1,g} \nonumber\\
    \forall t=2,...,T-1, g \in G^T: M_g^{on} < 2
\end{align}

\subsection{Offer Strategies}
In order for generators to bid strategically, the no load costs, startup costs, and variable cost in each offer step are further indexed by $t$. Generators can choose between three different offer strategies:
\begin{itemize}
    \item \textbf{Economic}: The generator submits its true costs and is eligible for make-whole payments.
    \item \textbf{Self-commit}: For each time period in which the generator wishes to self-commit, it submits an offer with zero startup and no-load costs, and no variable costs up to $P^{min}$. Variable costs beyond the minimum generation level are submitted economically. The generator is not eligible for make-whole payments.
    \item \textbf{Self-schedule}: For each time period in which the generator wishes to self-schedule, it submits an offer with zero startup and no-load costs, and no variable costs up to the desired dispatch quantity $p_{tg}^*$. Variable costs above $P^min$ are retained as economic offers, i.e., a generator is willing to be scheduled above its preferred schedule but submits an offer at its true cost. If a generator wishes to be scheduled for $ p_{tg}^* < \sum_{s=1,...,S^*} \rho_{tgs}$, then the variable cost of generation $C_{gS^{*}}$ is submitted for offers between $p_{tg}^*$ and $\sum_{s=1,...,S^*} \rho_{tgs}$. The generator is not eligible for make-whole payments.
\end{itemize}

We refer to strategic bids collectively as self-schedules or self-commits. This method of defining self-schedules and self-commits guarantees that no matter how many generators bid strategically, the system operator's problem is still feasible. Note that the only side payments generators can receive are make-whole payments in the case of an economic bid. Lost opportunity costs are never explicitly compensated, and doing so may create a revenue adequacy problem for the system operator. If lost opportunity costs were paid in full, under convex hull pricing there would be an incentive to submit arbitrarily large bids of zero price that must be accepted entirely if at all \cite{van_vyve_linear_2011}. 

\subsection{Greedy Algorithm}

The generators' problem of choosing what offer strategy to bid is a multi-armed bandit problem. In a multi-armed bandit problem, a set of discrete choices results in uncertain, random payoffs. The bandit (gambler) seeks to find a strategy that maximizes payoff. Each generator must determine how to bid based on the profitability of each strategy determined in previous outcomes, which depends on the offer strategies of other generators. One method to solve the multi-armed bandit problem is via the greedy reinforcement learning algorithm \cite{Kuleshov2014,gummadi_mean_2013}. Drawing from a history of prior outcomes, the greedy algorithm chooses the best strategy in expectation with probability $\alpha$ and chooses a strategy at random with probability $1-\alpha$. 

In the first iteration of the simulation, all generators bid economically, representing the competitive market solution. Afterwards the generators explore other strategies randomly with probability $1-alpha$. If the expected payoff is equivalent between a strategic bid and an economic bid, the generator defaults to bidding economically. 

\subsection{Exponential Smoothing}

The expected profit of a strategy is calculated via exponential smoothing with parameter $\eta$. Let $x$ be a vector of values to be smoothed indexed by $t$ of length $T$. The expected value of $x$ is $s_T$, where $s_t$ is calculated as:

\begin{align}
    s_1 &= x_1 \\
    s_t &= \eta x_t + (1-\eta)s_{t-1} \quad \forall t \geq 2
\end{align}

In order to determine when in a multi-period optimization horizon to self-commit or self-schedule, a generator must determine an expected price stream for a given strategy. To calculate the expected prices, the price $\lambda_t$ in a given period is exponentially smoothed across all iterations in which the generator chose the given strategy.

\subsection{Adverse Bidding Test}

We define a strategic bid as adverse if bidding strategically increases the expected profits of a generator. For an adverse bidder, either $\overline{X}_{selfcomm} - \overline{X}_{eco} > 0$ or $\overline{X}_{selfsched} - \overline{X}_{eco} > 0$. Note that an adverse bid may result in an increase in total production costs and thus a decrease in market efficiency, but it may also represent a transfer of profits among generators without impacting the total social surplus achieved. The payoffs a generator makes are influenced not only by its bidding decision but the bidding decisions of the other generators. A generator may learn to bid strategically based on profits achieved due to the bidding strategy of others. To determine if a strategic bidder is actually able to increase its profits via its own offer strategy, we use the an unequal variance two-sample t-test, also known as Welch's t-test.

Welch's t-test tests the null hypothesis that two sets of samples come from distributions with equal means against the alternative hypothesis that the distributions have different means. Like Student's t-test, it assumes the sample means being compared are normally distributed, but unlike Student's t-test, it does not assume that the populations have equal variances. It is more reliable than Student's t-test when the samples have unequal variances and unequal sample sizes \cite{welch1947generalization}.

The test statistic $t$ is defined as:

\begin{align}
    t =& \frac{\Delta \overline{X}}{s_{\Delta \overline{X}}} = \frac{\overline{X}_1 - \overline{X}_2}{\sqrt{s^2_{\overline{X}_1} +s^2_{\overline{X}_2}}} \\
    s_{\overline{X}_i} =& \frac{s_i}{\sqrt{N_i}}
\end{align}

where $\overline{X}_i$ and $s_{\overline{X}_i}$ is the sample mean and its standard error, $s_i$ is the corrected sample standard deviation, and $N_i$ is the sample size. In the analysis that follows, we use a p-value of 0.05.

\section{Illustrative Test Case}\label{sec:stylizedtestcase}

Authors in \cite{eldridge_pricing_2018,eldridge_algorithms_2020} propose a stylized test case in which the Nash equilibria strategies can be found analytically. We use this test case for three scenarios: a single-period case $|T|=1$ with demand profile $D_1$, a multi-period case with $|T|=10$ and constant demand profile $D_1$, and a multi-period cast $|T|=1$ with fluctuating demand profile $D_2$. The case consists of 3 types of generators with characteristics given in Table \ref{tab:BEexample}.

\begin{table}[htb!]
    \centering
    \caption{Illustrative test case generator characteristics}
    \begin{tabular}{ cccc } 
     \toprule
     Gen. $i \in \{1,...,5\}$ & $P^{min}_{hi}$ (MW) & $P^{max}_{hi}$ (MW) & $C_{hi}$ (\$/MWh) \\
     \midrule
     $GEN1_i$ & 25 & 25 & 15\\ 
     $GEN2_i$ & 0 & 25 & 10\\ 
     $GEN3_i$ & 0 & 25 & 25\\ 
    \bottomrule
    \end{tabular}
    \label{tab:BEexample}
\end{table}

Generators types GEN2 and GEN3 are convex with only a marginal cost, while type GEN1 is block-loaded, making the optimal value function of this system non-convex. Let the demand level be $225 + \epsilon $ MW, where $\epsilon > 0$ and negligibly small. 

Under the FCP model, the price is \$25 for the socially optimal solutions in which any 4 GEN1s are committed as well as for any integer solution in which less than 4 of the 5 GEN1s self-commit. If all 5 GEN1s self-commit, the price decreases to \$10, and the total actual production costs to serve load increase, leading to a decrease in market efficiency. If 4 of the 5 GEN1s are committed, the $5^{th}$ has a LOC of $-\$250$, as it would prefer to be committed given the price of $\$25$.\footnote{The convention used here is to display LOC as the perceived profit or loss. A perceived loss is negative. A perceived profit is possible if a side payment was given in excess of perceived losses, e.g., if CHP were calculated over a different time period than MWP.} If all GEN1s self-commit and the price drops to \$10, then each suffers a loss, with a payoff of -\$125. Let $\gamma_{hi}$ be the probability that a generator self-commits. Assuming there is no collusion, the mixed strategy Nash equilibrium can be found analytically to be $\gamma_{1i} = 0.831$ $\forall i \in \{1,...,5\}$ \cite{eldridge_pricing_2018}. There are also five fixed asymmetric strategies in which 4 of the 5 generators self-commit and one does not. In contrast, under the CHP model, the price is \$15, and the GEN1s have no incentive to self-schedule, as their profits are always 0.

We replicate the above example and show that the simulations converge on one of the Nash equilibria with fixed asymmetric strategies. The single-period scenario $|T|=1$ has demand profile $D_1$. Removing all constraints in \eqref{eq:SS:UC} for which $t>1$, we simulate the market using greedy $\alpha = 0.9$
and exponential smoothing $\eta = 0.05$ over 2000 iterations.

\begin{figure}[tbh]
    \centering
    \includegraphics[width=.8\textwidth]{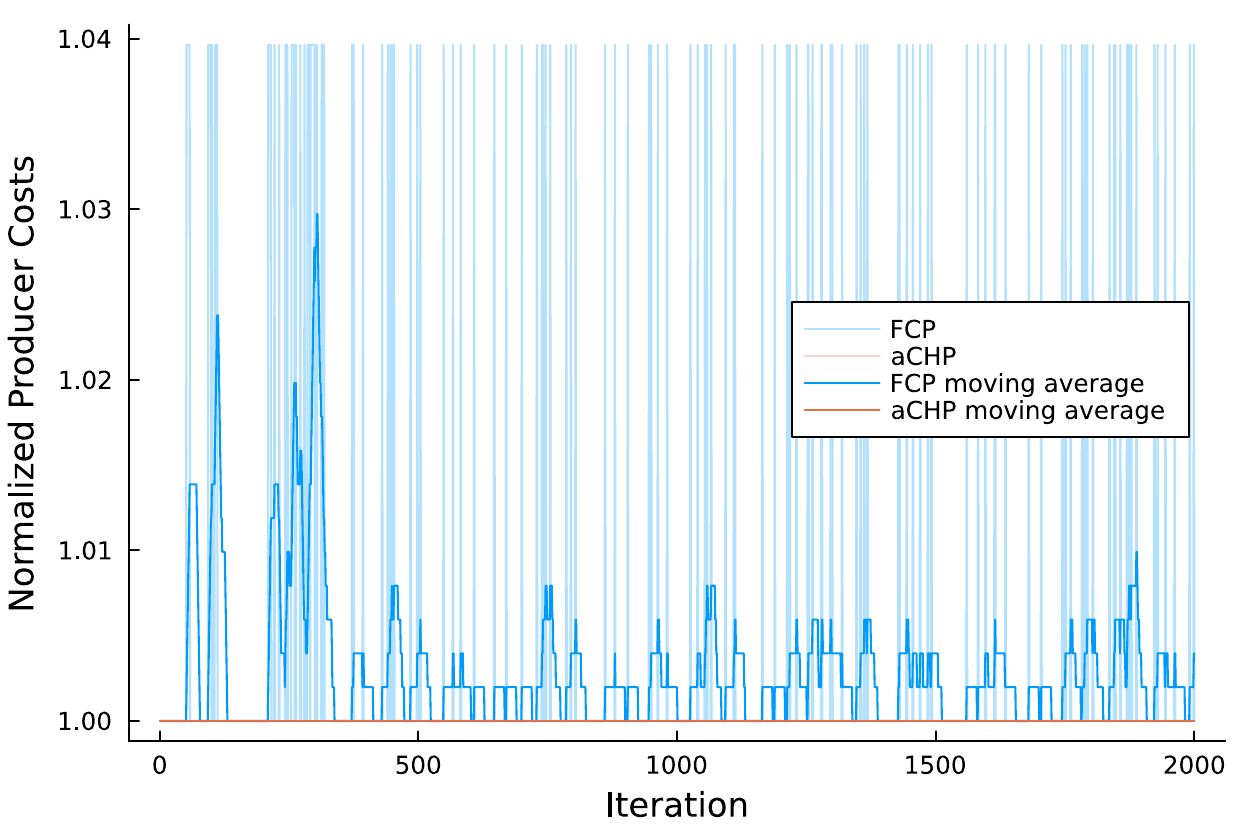}
    \caption{$|T|=1$, $D_1$ Total actual production cost normalized by production cost at the competitive solution in which all generators bid economically. Higher producer costs indicate the system operator selected a suboptimal solution due to strategic bids.}
    \label{fig:toyBE_1h_producercosts}
\end{figure}

\begin{figure}[tbh]
    \centering
    \includegraphics[width=.8\textwidth]{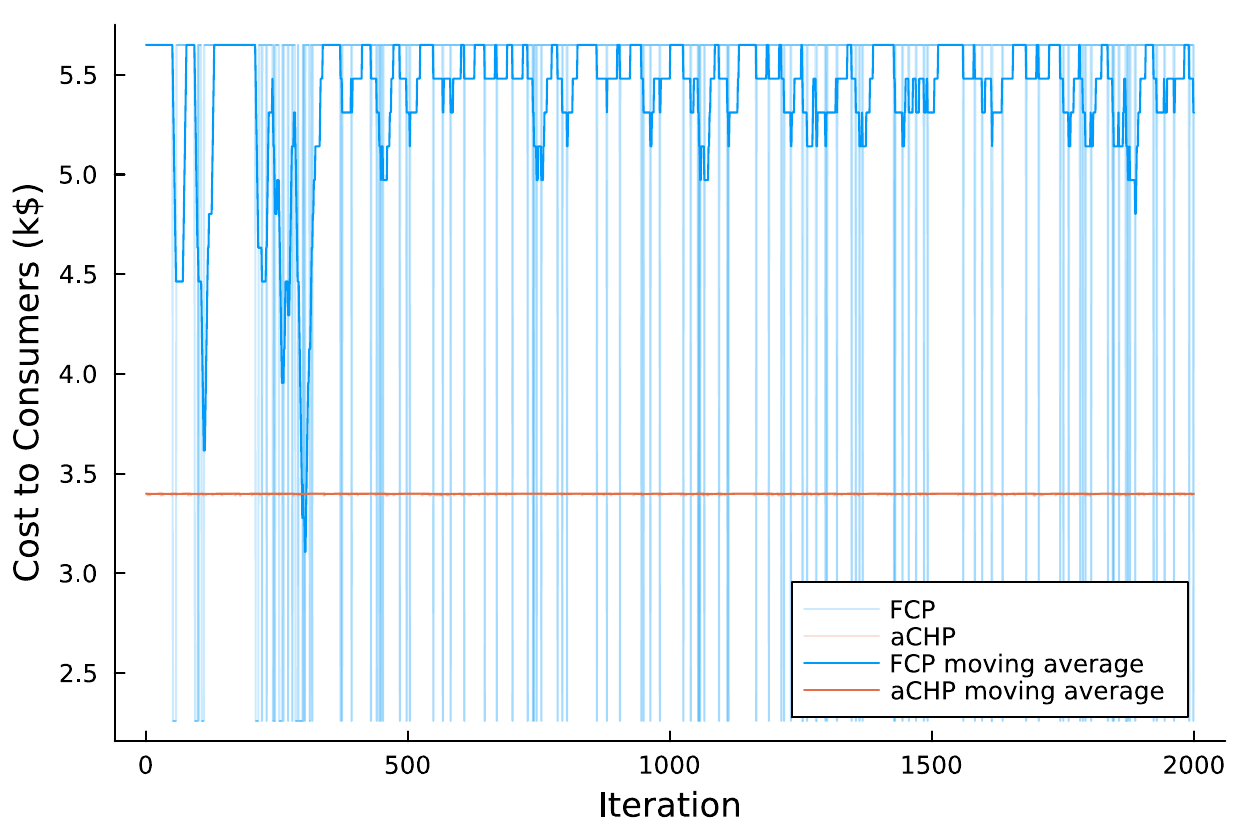}
    \caption{$|T|=1$, $D_1$. Cost to consumers for each pricing model over iterations.}
    \label{fig:toyBE_1h_consumercosts}
\end{figure}

Recall that the first iteration is the competitive solution in which all generators bid economically. Figure \ref{fig:toyBE_1h_producercosts} shows that actual total producer costs do not increase above the competitive solution once strategic bidding is allowed. The system operator never selects a suboptimal solution due to the strategic bidding. The generators are also not able to increase total producer profits by strategic bidding. Since demand is inelastic and all demand is served, no increase in producer profits implies no increase in the cost to consumers, shown in Figure \ref{fig:toyBE_1h_consumercosts}. When all GEN1s self-commit/schedule, the price drops to \$10 and no make-whole payments are required, leading to lower costs to consumers and lower producer profits in these iterations.

\begin{figure}[hbt]
    \centering
    \includegraphics[width=.8\textwidth]{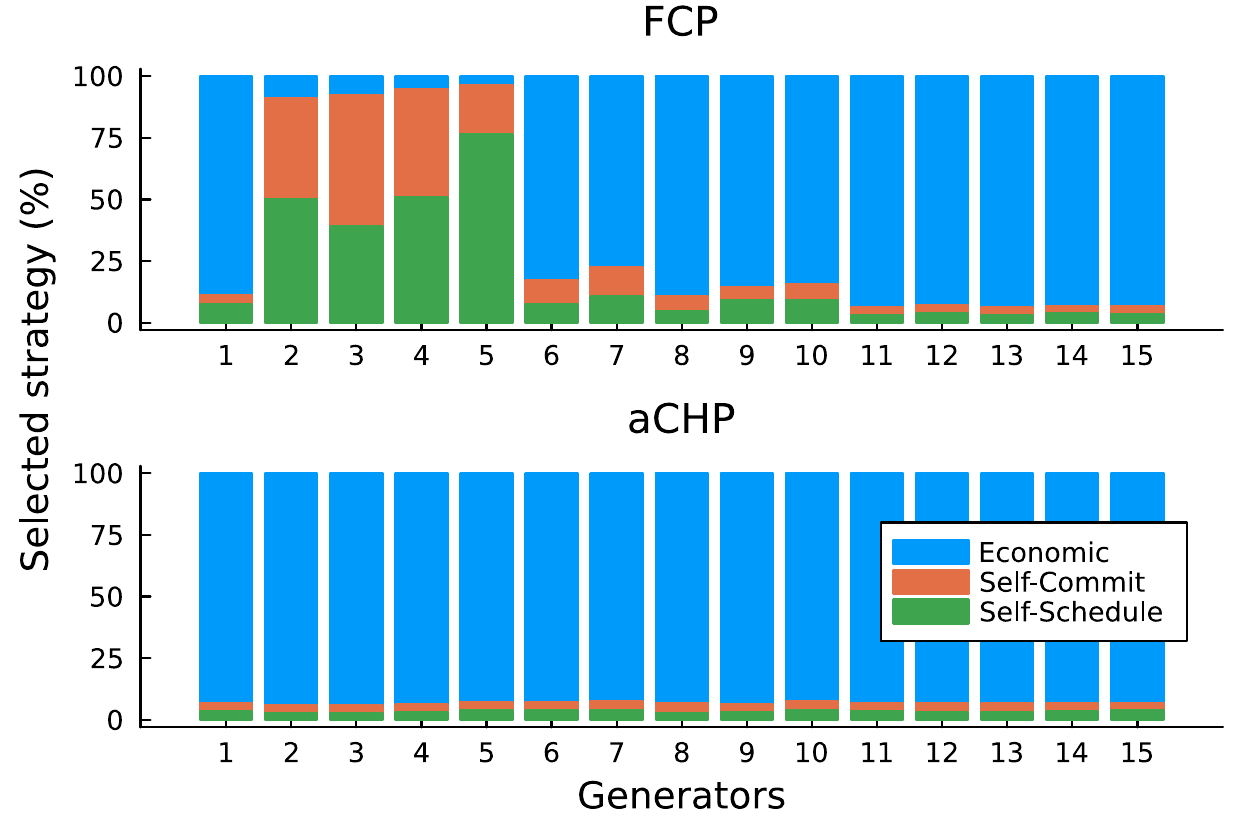}
    \caption{$|T|=1$, $D_1$. Percent of iterations that a generator chose each offer strategy.}
    \label{fig:toyBE_1h_selectedstrat}
\end{figure}

The number of times each generator selects each offer strategy over the simulation period is shown in Figure \ref{fig:toyBE_1h_selectedstrat}. Under the FCP model, 4 of the 5 GEN1s learn to self-commit (or, equivalently, self-schedule, as the unit is block-loaded) while the GEN1$_1$ learns to bid economically. GEN1s collectively self-commit/schedule 78.4\% of the time over the final 1000 iterations. A number of GEN2s self-commit or self-schedule more than the random exploratory probability $(1-\alpha)/3$. However, this does not mean that they learn to strategically bid because they are necessarily able to influence the price. A GEN2 may bid strategically and see a bigger payoff (or bid economically and see a smaller payoff) incidentally because of the behavior of GEN1s in that iteration. However, no GEN2 obtains mean profits self-committing or self-scheduling that are statistically significantly greater than the mean profits from bidding economically.


\begin{figure}[tbh]
    \centering
    \includegraphics[width=.8\textwidth]{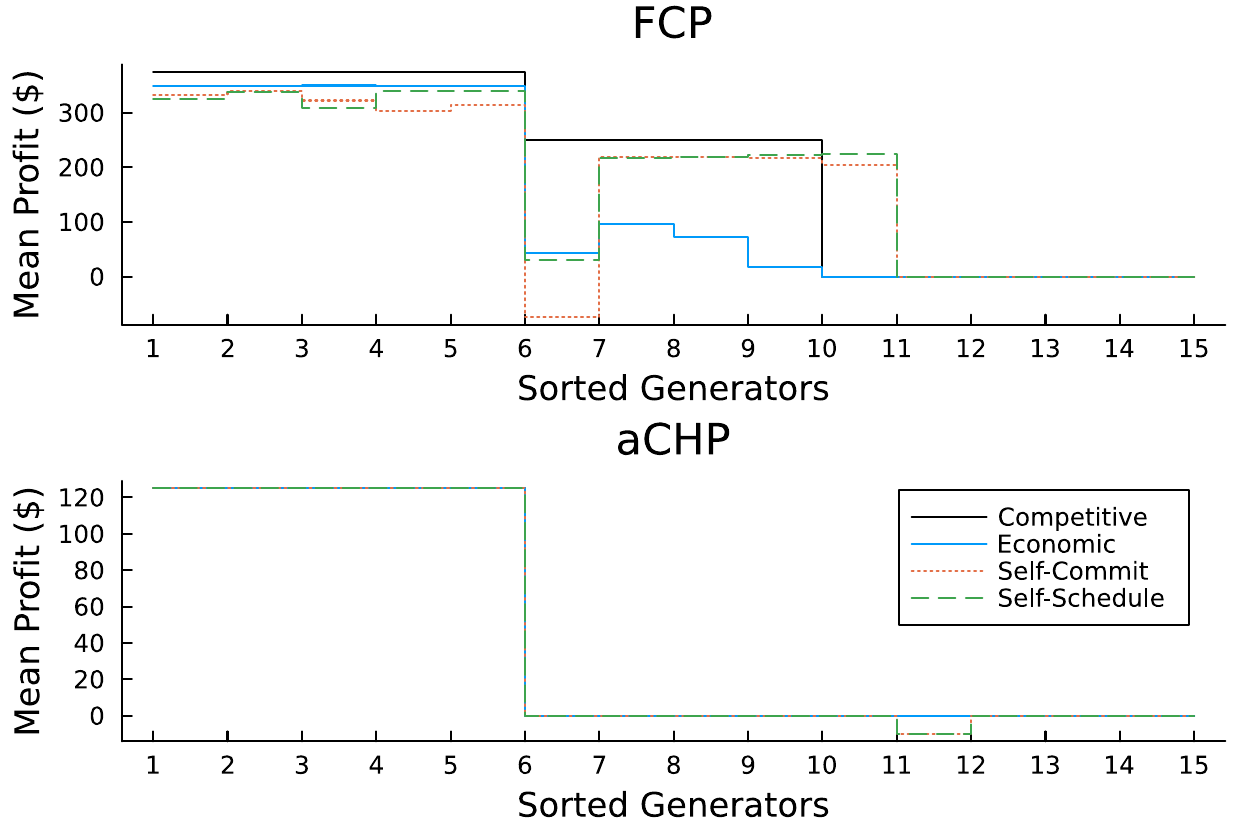}
    \caption{$|T|=1$, $D_1$ Profit duration curve. Generators are sorted by profit achieved in the competitive outcome in which all generators submit economic bids. The mean profit achieved for each strategy in simulation is also shown.}
    \label{fig:toyBE_1h_profitduration}
\end{figure}

A generator is said to be an adverse strategic bidder if it profits in expectation from bidding strategically rather than bidding economically in a statistically significant manner. Either $\overline{X}_{selfcomm} - \overline{X}_{eco} > 0$ or $\overline{X}_{selfsched} - \overline{X}_{eco} > 0$ and the p-value from the corresponding Welch's t-test is $> 0.05$. Table \ref{tab:AdverseStratBidsFCP} and Table \ref{tab:AdverseStratBidsaCHP} show the number of adverse strategic bidders under the FCP and aCHP models. For the FCP model, 4 of the GEN1s are statistically significant adverse strategic bidders, while 0 of the GEN2s are. Under the aCHP model, no generators pass the statistical significance test.

\begin{table}[H]
    \centering
    \caption{Adverse Strategic Bids (FCP)}
    \begin{tabular}{ cccc } 
     \toprule
     &\multicolumn{3}{c}{Number}\\
       &\multicolumn{3}{c}{$\overline{X}_{eco} < (\overline{X}_{selfsched}$ OR $ \overline{X}_{selfcomm}) $} and $p<0.05$\\
      Gen. $i \in \{1,...,5\}$&$|T|=1$ , $D_1$&$|T|=10$, $D_1$ &$|T|=10$, $D_2$\\
     \midrule
     $GEN1_i$ & 4 & 4 & 5\\ 
     $GEN2_i$ & 0 & 0 & 3\\ 
     $GEN3_i$ & 0 & 0 & 0\\ 
    \bottomrule
    \end{tabular}
    \label{tab:AdverseStratBidsFCP}
    \vspace{.5cm}
    \caption{Adverse Strategic Bids (aCHP)}
    \begin{tabular}{ cccc } 
     \toprule
     &\multicolumn{3}{c}{Number}\\
       &\multicolumn{3}{c}{$\overline{X}_{eco} < (\overline{X}_{selfsched}$ OR $ \overline{X}_{selfcomm}) $} and $p<0.05$\\
      Gen. $i \in \{1,...,5\}$&$|T|=1$, $D_1$ &$|T|=10$, $D_1$ &$|T|=10$, $D_2$\\
     \midrule
     $GEN1_i$ & 0 & 0 & 0\\ 
     $GEN2_i$ & 0 & 0 & 3\\ 
     $GEN3_i$ & 0 & 0 & 0\\ 
    \bottomrule
    \end{tabular}
    \label{tab:AdverseStratBidsaCHP}
\end{table}


Excess profit is defined as the difference for strategic generators between the mean profit for the strategic bidding strategy with the highest payoff (either $\overline{X}_{selfcomm}$ or $\overline{X}_{selfsched}$) and $\overline{X}_{eco}$, the mean profit when bidding economically. The total excess profit among all adverse strategic bidders is shown as a percentage of the total producer profits in the competitive solution in which all generators bid economically for each pricing model in Table \ref{tab:AdverseStratBidsPayoffs}. Note that it is typically not possible for generators to realize their excess profit simultaneously, but this figure gives a sense of the magnitude of the profit opportunity for strategic bidders in the market. Excess profits under FCP are 24.4\%, while there are no excess profits under aCHP.

\begin{table}[H]
    \centering
    \caption{Adverse Strategic Bids Payoffs}
    \begin{tabular}{ cccc } 
     \toprule
       &\multicolumn{3}{c}{ Total excess profit (\% competitive profits)}\\
      Pricing Model&$|T|=1$ , $D_1$&$|T|=10$, $D_1$ &$|T|=10$, $D_2$\\
     \midrule
     FCP & 24.4\% & 20.9\% & 13.6\%\\ 
     aCHP & 0\% & 0\% & 0.08\%\\ 
    \bottomrule
    \end{tabular}
    \label{tab:AdverseStratBidsPayoffs}
\end{table}

Figure \ref{fig:toyBE_1h_profitduration} shows the profit duration curve for generators at the competitive solution and the mean profits made in simulation with each bidding strategy. In the competitive solution, all GEN2s are committed and make the highest profits \$(25-10)($D_1 + \epsilon$), while 4 of the GEN1s are committed and make a profit \$(25-15)($D_1 + \epsilon$), with 1 GEN1 not committed and making no profit. One GEN3 is committed to clear demand $\epsilon$ but makes no profit. Under the aCHP model, all generators have the same payoff as under the competitive solution regardless of bidding strategy, with the exception of the marginal GEN3 that discovers self-committing or self-scheduling will entail a loss. For FCP, the GEN2s do best by bidding economically, but suffer a loss in expectation compared to the competitive solution due to GEN1 strategic behavior sometimes lowering the price to \$10 from \$25. A profit transfer takes place among the GEN1s, in which a generator that is committed in the competitive solution and profits is shut out by the other four generators' strategic behavior.

\begin{figure}[tbh]
    \centering
    \includegraphics[width=.8\textwidth]{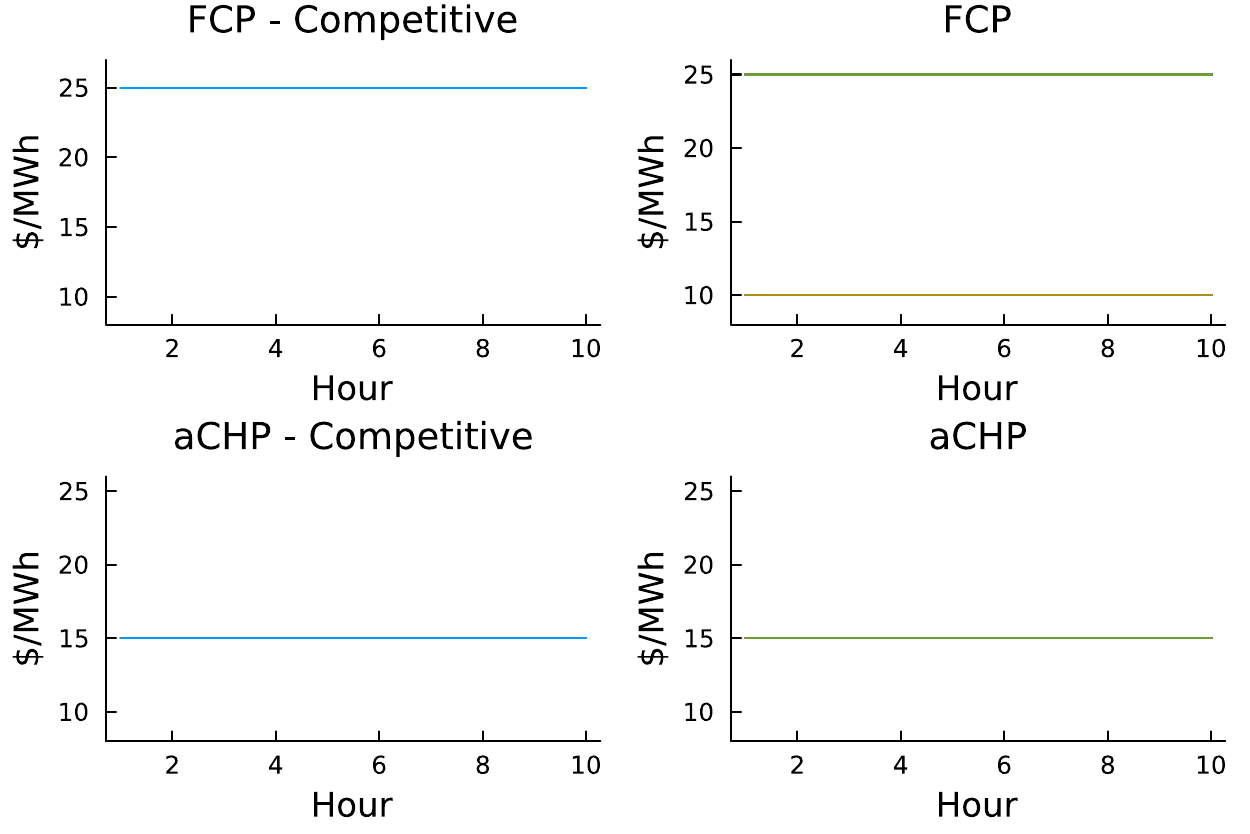}
    \caption{$|T|=10$, $D_1$. Prices attained under the competitive solution in which all generators bid economically and prices attained over all iterations of strategic bidding.}
    \label{fig:toyBE_10h_toyprices}
\end{figure}

 Next, we expand this market into a multi-period market, assuming no binding ramping constraints. The market has constant demand $D_1$ of 225 + $\epsilon$, where $\epsilon = 1$ MW, yielding the same price possibilities as the prior example. The competitive prices and the prices found in simulation with strategic bidding are shown in Figure \ref{fig:toyBE_10h_toyprices}. 
 
\begin{figure}[tbh]
    \centering
    \includegraphics[width=.8\textwidth]{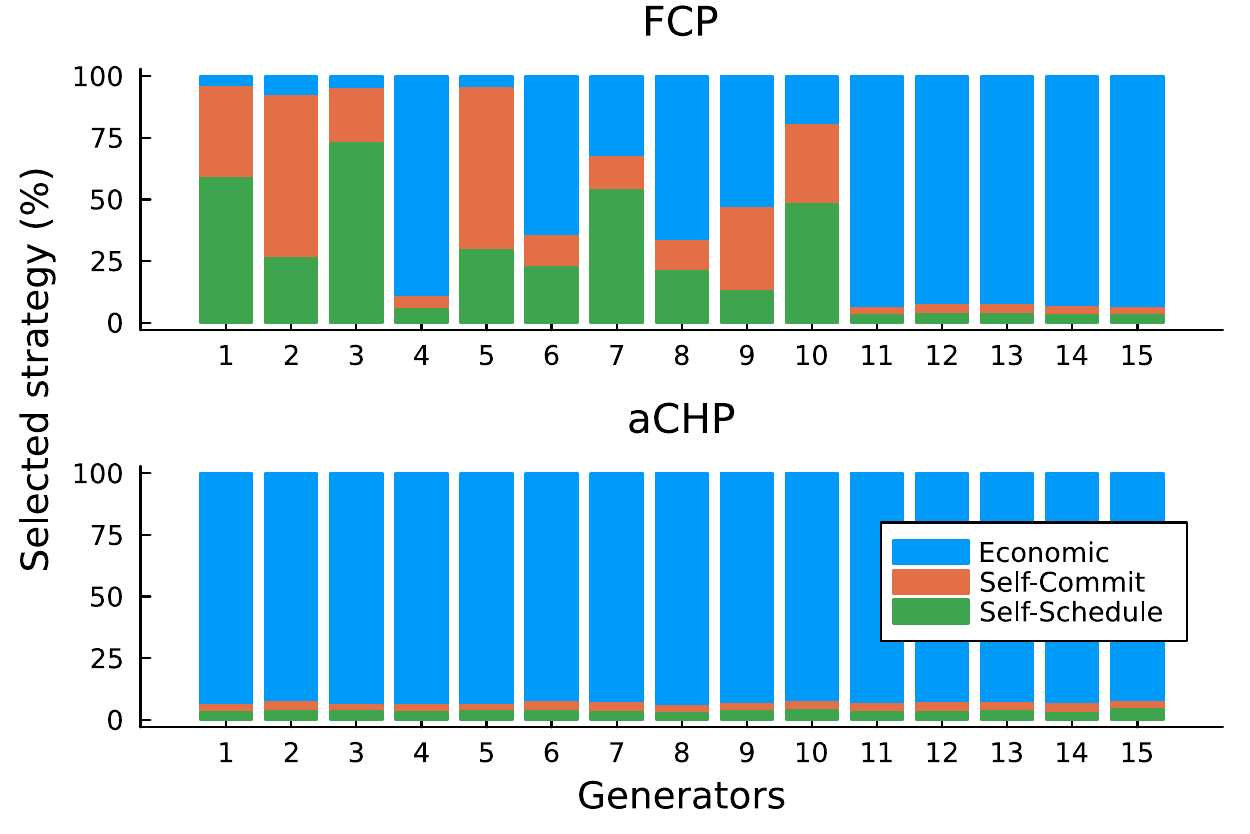}
    \caption{$|T|=10$, $D_1$. Percent of iterations that a generator chose each offer strategy.}
    \label{fig:toyBE_10h_selectedstrat}
\end{figure}

 While the learning behavior shown in Figure \ref{fig:toyBE_10h_selectedstrat} appears different than when $|T|=1$, Tables \ref{tab:AdverseStratBidsFCP} and \ref{tab:AdverseStratBidsaCHP} show that the number of statistically significant adverse bidders in each pricing model is equivalent. A different asymmetric Nash equilibrium is found in which GEN1$_4$ bids economically and all others bid strategically. GENS1s collectively self-commit/schedule 78.5\% of the time over the final 1000 iterations and excess profits shown in Table \ref{tab:AdverseStratBidsPayoffs} are similar to $|T|=1$.

\begin{figure}[tbh]
    \centering
    \includegraphics[width=.8\textwidth]{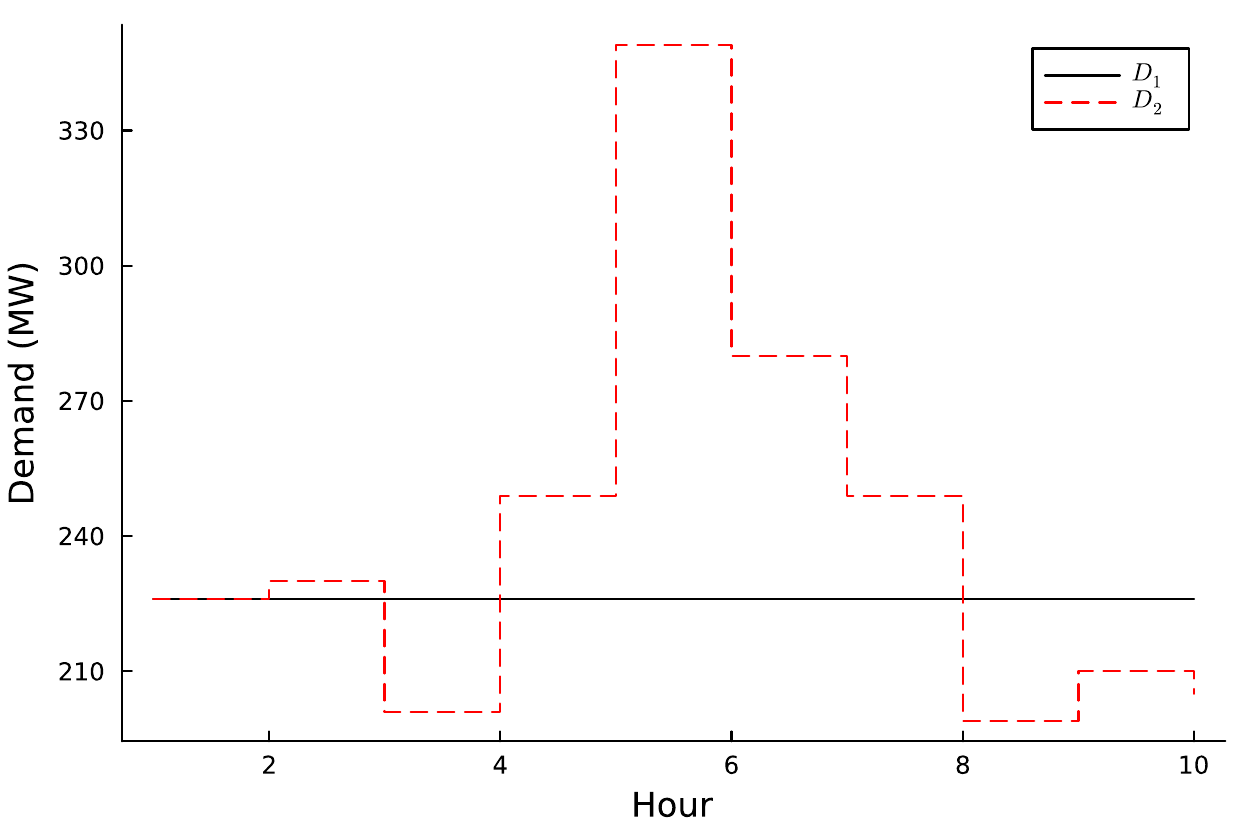}
    \caption{Demand profiles.}
    \label{fig:toyBE_10hvary_toydemand}
\end{figure}

\begin{figure}[tbh]
    \centering
    \includegraphics[width=.8\textwidth]{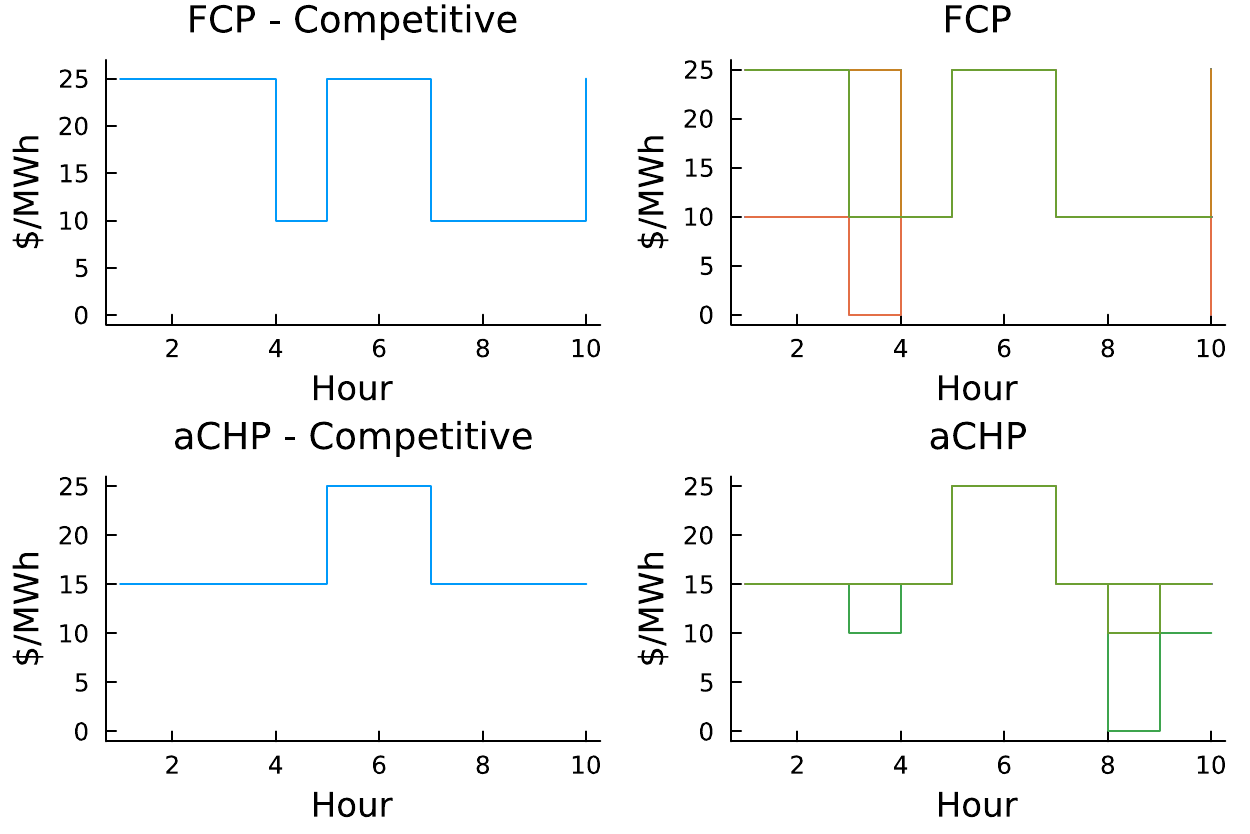}
    \caption{$|T|=10$, $D_2$. Prices attained under the competitive solution in which all generators bid economically and prices attained over all iterations of strategic bidding.}
    \label{fig:toyBE_10hvary_toyprices}
\end{figure}

\begin{figure}[tbh]
    \centering
    \includegraphics[width=.8\textwidth]{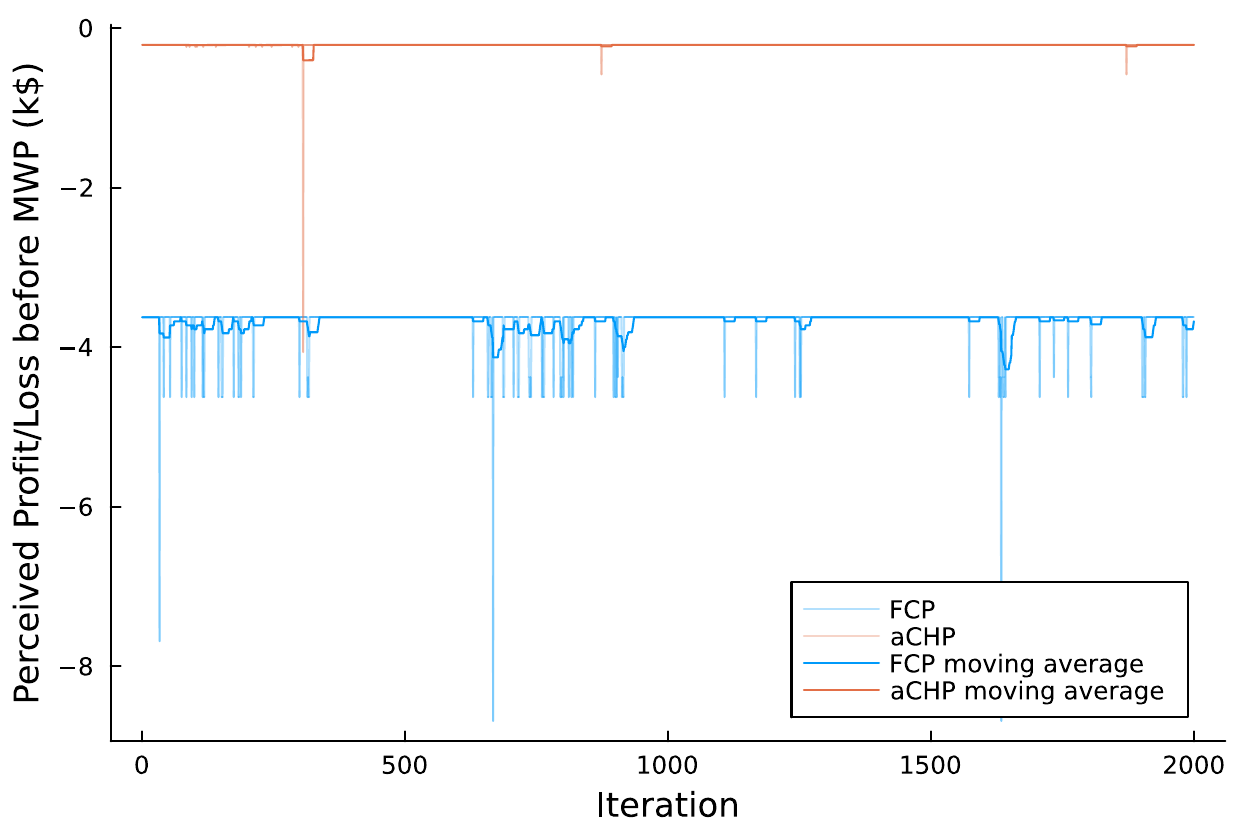}
    \caption{$|T|=10$, $D_2$. Lost opportunity cost displayed as perceived profit or loss before MWP.}
    \label{fig:toyBE_10hvary_LOC}
\end{figure}

 Next we vary the demand, with demand profile $D_2$ shown in Figure \ref{fig:toyBE_10hvary_toydemand} and $|T|=10$. This yields the competitive and strategic bidding prices shown in Figure \ref{fig:toyBE_10hvary_toyprices}. The lost opportunity costs for aCHP are lower than FCP, as shown in Figure \ref{fig:toyBE_10hvary_LOC}. 
 
\begin{figure}[tbh]
    \centering
    \includegraphics[width=.8\textwidth]{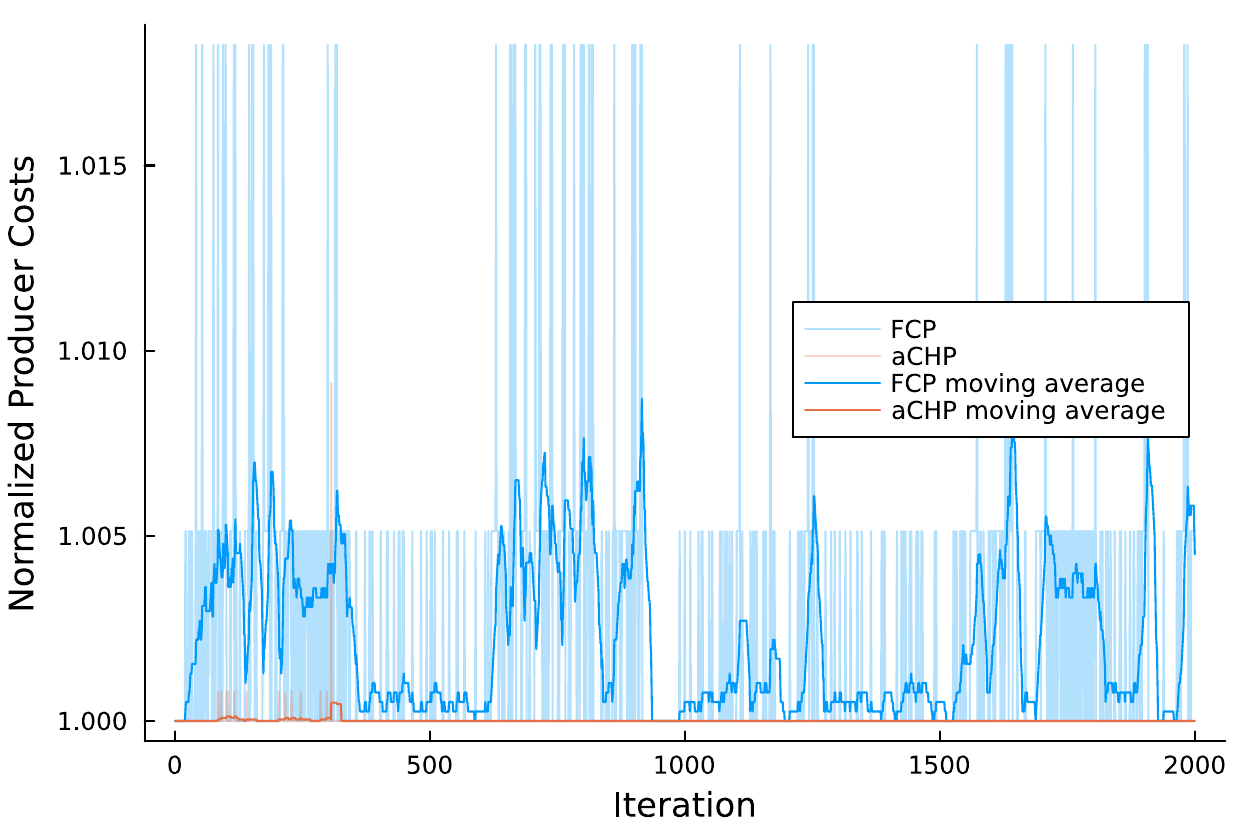}
    \caption{$|T|=10$, $D_2$ Total actual production cost normalized by production cost at the competitive solution in which all generators bid economically.}
    \label{fig:toyBE_10hvary_producercosts}
\end{figure}

 The total producer costs normalized by the costs in the competitive solution are shown in Figure \ref{fig:toyBE_10hvary_producercosts}. Strategic bidding with FCP can increase actual producer costs by over 1.5\%, but on average only increases costs slightly ($<0.5\%$). 

\begin{figure}[tbh]
    \centering
    \includegraphics[width=.8\textwidth]{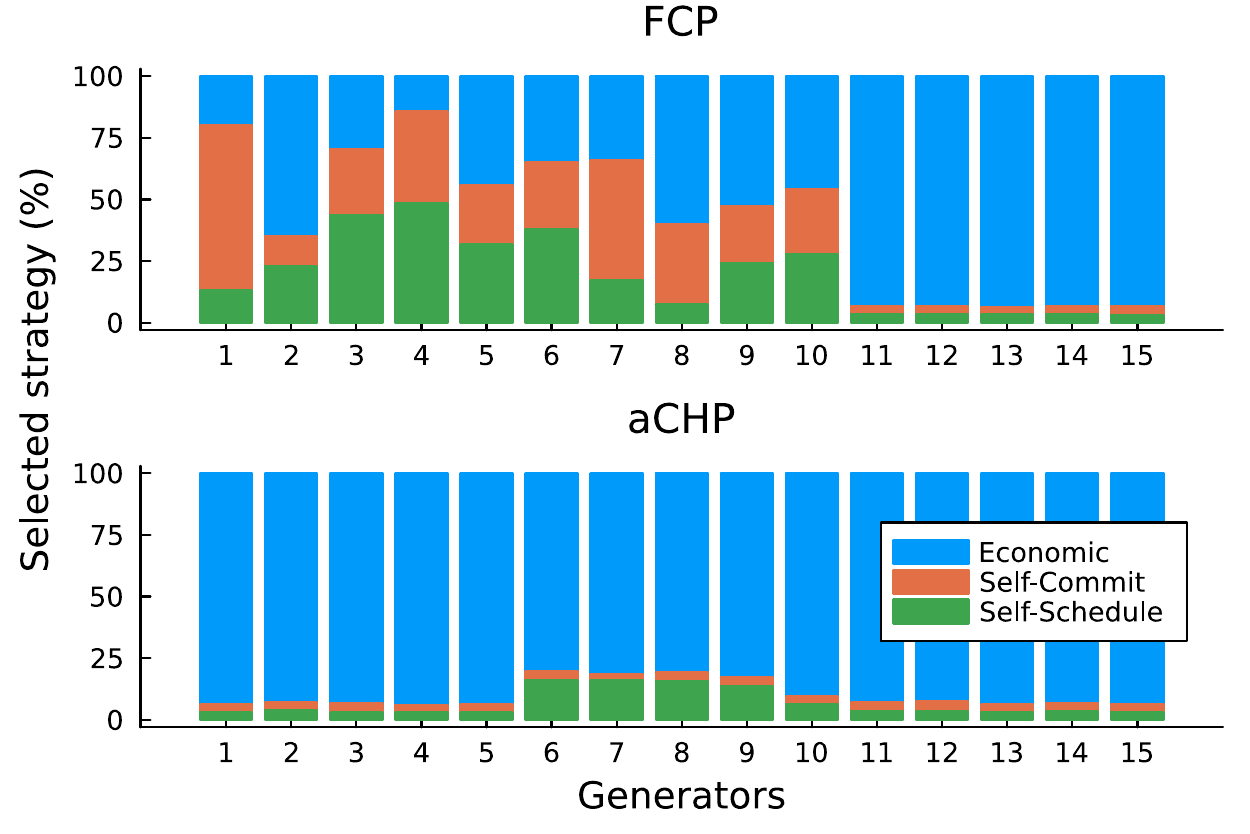}
    \caption{$|T|=10$, $D_2$. Percent of iterations that a generator chose each offer strategy.}
    \label{fig:toyBE_10hvary_selectedstrat}
\end{figure}

\begin{figure}[tbh]
    \centering
    \includegraphics[width=.8\textwidth]{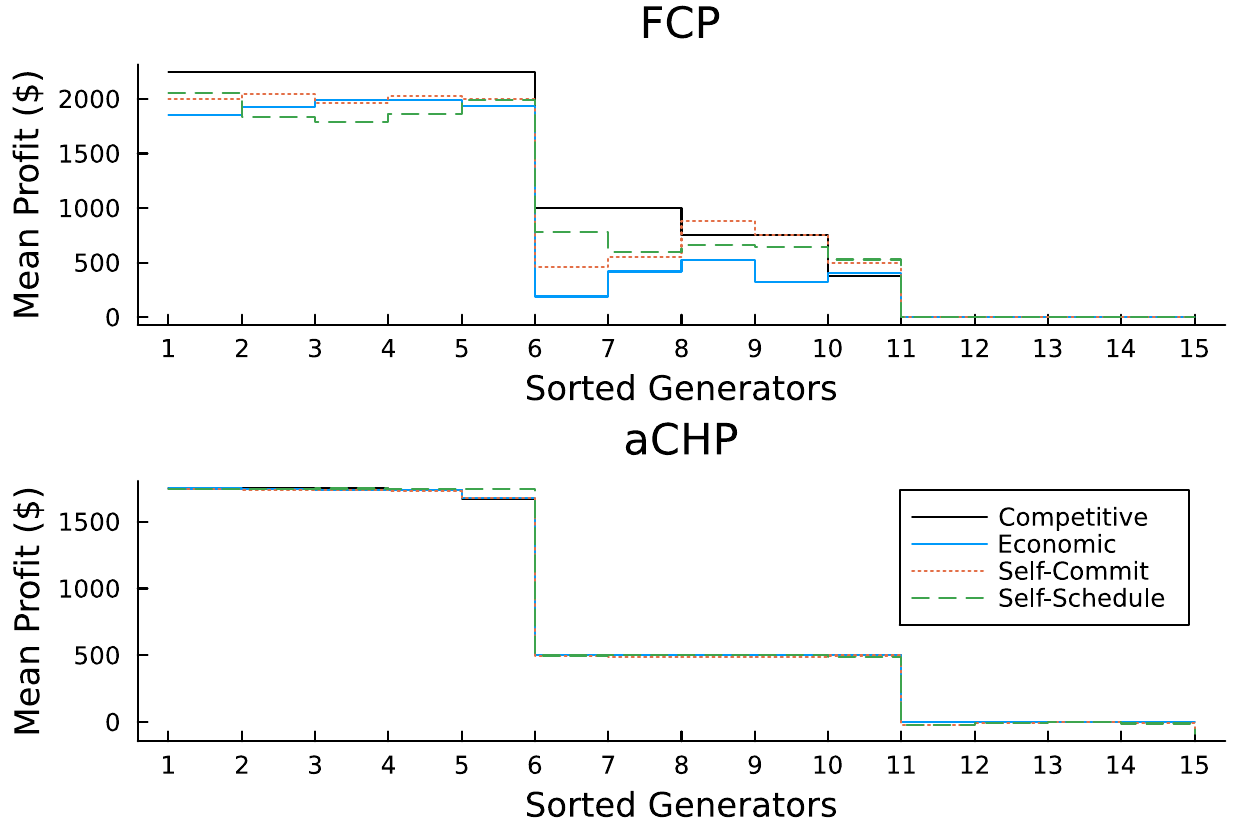}
    \caption{$|T|=10$, $D_2$ Profit duration curve. Generators are sorted by profit achieved in the competitive outcome in which all generators submit economic bids.}
    \label{fig:toyBE_10hvary_profitduration}
\end{figure}

Figure \ref{fig:toyBE_10hvary_selectedstrat} shows the percent of iterations each generator chose each strategy. The number of adverse bidders is again shown in Tables \ref{tab:AdverseStratBidsFCP} and \ref{tab:AdverseStratBidsaCHP}. Under FCP, GEN1s all are adverse strategic bidders, collectively self-committing/scheduling 64.7\% of the time over the final 1000 iterations. GEN2s collectively self-schedule or self-commit 62.0\% of the final 1000 iterations. All 5 GEN1s are statistically significant adverse bidders, but only 3 of the 5 GEN2s are. Self-committing is profitable for 2 GEN2s but self-scheduling is profitable for 3. Under aCHP, 3 GEN2s also learn to profitably bid strategically by self-scheduling, but the payoff is very small. The total excess profit as a percentage of the competitive profits for FCP is 13.6\%, while it is only 0.08\% for aCHP, shown in Table \ref{tab:AdverseStratBidsPayoffs}. The profit duration curve at the competitive solution is shown in Figure \ref{fig:toyBE_10hvary_profitduration}. While the mean profits vary little under aCHP no matter the offer strategy, the mean profits attained for each strategy under FCP vary considerably. 

\section{Large-Scale Test Case}

\subsection{Data}

Data for the large-scale test system come from a benchmark library curated and maintained by the IEEE PES Task Force on Benchmarks for Validation of Emerging Power System Algorithms
\cite{ieee_pes_power_grid_benchmarks_power_2019}. We simulate several days from the FERC test cases \cite{krall2012rto,knueven2020mixed}. Generator data are based on the publicly available unit commitment test instance from the Federal Energy Regulatory Commission (FERC) consisting of approximately 1000 generators with load and wind data based on publicly available data from PJM. Each case includes an aggregated variable renewable energy (VRE) generator. This is a wind profile that is scaled to be 2\% of annual load in the low wind scenarios and 30\% in the high wind scenarios.

Generators have up to 12 offer steps $S$, and marginal costs in each offer step were calculated from cumulative costs in each offer step. Block loaded units are assumed to have no no-load costs. The variable cost from 0 to $P^{min}$ is assumed to be the same variable cost as in the first offer step above $P^{min}$, with the remainder as no-load cost. If this is infeasible (e.g., if the no-load cost would be negative), there is assumed to be no no-load cost, and the variable cost between 0 and $P^{min}$ is the cumulative cost to produce at $P^{min}$ divided by $P^{min}$.

The three cases considered are shown in Table \ref{tab:FERCdata}. FERC1 and FERC3 are high wind cases, while FERC 2 is the low wind version of FERC1. While no resource mixes in these test cases are adapted in the long-run to the demand profiles, we expect FERC2 is especially poorly adapted because of the large increase in wind capacity, i.e., many thermal resources in this system ought to be incentivized to exit the market. FERC1 and FERC2 are a summer day and FERC3 is a winter day. Table \ref{tab:FERCdata} lists the maximum aggregate wind generation and maximum demand over the 24-hour period.

\begin{table}[H]
    \centering
    \caption{FERC Test Cases}
    \begin{tabular}{ rlll } 
     \toprule
     &FERC1 & FERC2 & FERC3\\
     \midrule
     Date & 2015-07-01 & 2015-07-01 & 2015-02-01 \\ 
     Max Demand & 112.6 GW & 112.6 GW & 103.7 GW \\
     Thermal Generators & 978 (177.5 GW) & 978 (177.5 GW)  & 934 (180.7 GW) \\ 
     VRE Generators & 1 (1.2 GW) & 1 (18.2 GW) & 1 (4.5 GW) \\ 
    \bottomrule
    \end{tabular}
    \label{tab:FERCdata}
\end{table}

\subsection{Results}

All optimizations are solved to a MIP gap of 0.01\% with a horizon of 24 hours with no look-ahead. Transmission constraints and reserve requirements are omitted. Since demand is considered as inelastic and all demand is cleared without any non-served energy, the MIP gap reflects total producer costs. Note the convention that if a generator is neutral between bidding strategically or bidding economically at a given iteration in the simulation period based on expected profit, the generator defaults to bidding economically. The first iteration is the competitive solution in which all generators bid economically. The simulation is run with greedy $\alpha = 0.9$ and exponential smoothing $\eta = 0.05$ over 1000 iterations.

 \begin{figure}[tb]
    \centering
    \includegraphics[width=.8\textwidth]{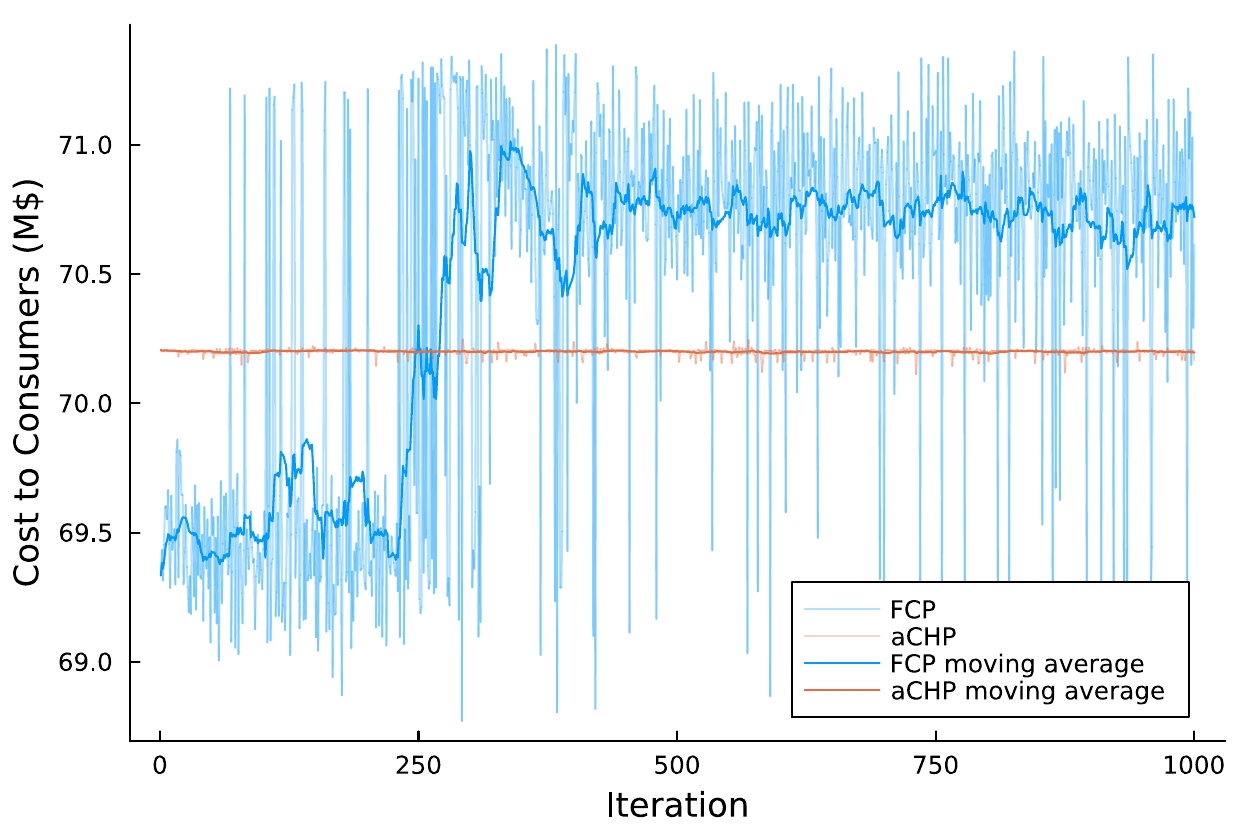}
    \caption{FERC1. Cost to consumers for each pricing model over iterations. Average costs to consumers under FCP rise during the simulation to be greater than costs to consumers under aCHP.}
    \label{fig:P1_consumercosts}
\end{figure}

\begin{figure}[tbh]
    \centering
    \includegraphics[width=.8\textwidth]{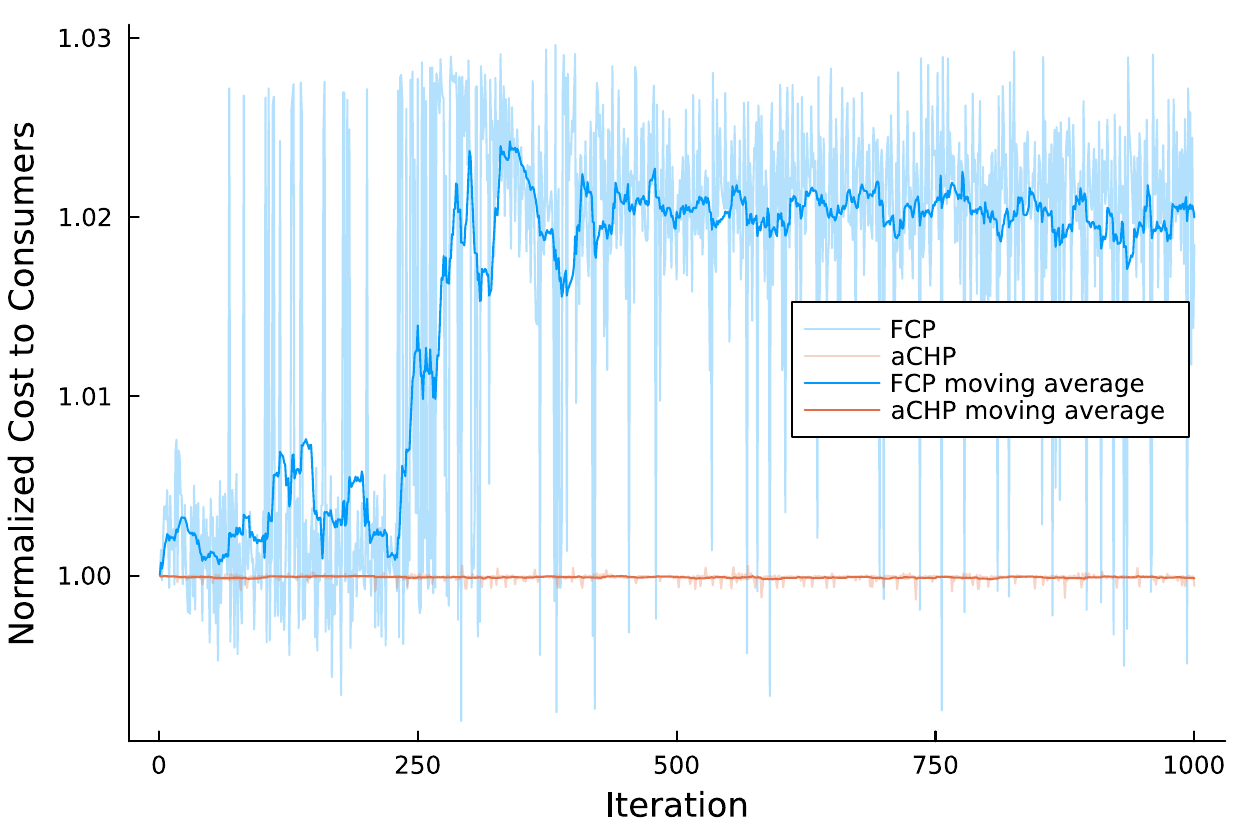}
    \caption{FERC1. Cost to consumers over iterations normalized by cost to consumers at the competitive solution in which all generators bid economically for each pricing model.}
    \label{fig:P1_normconsumercosts}
\end{figure}

\begin{figure}[tbh]
    \centering
    \includegraphics[width=.8\textwidth]{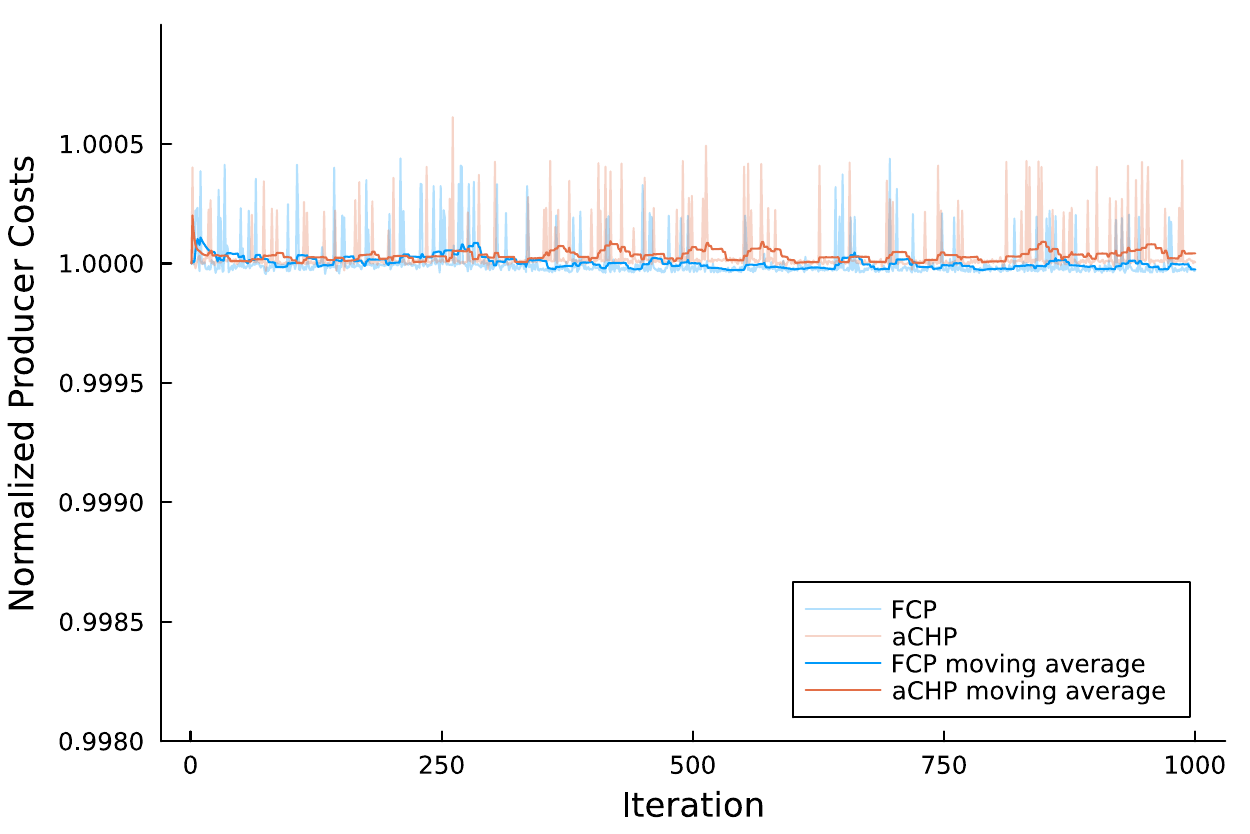}
    \caption{FERC1. Total actual production cost normalized by production cost at the competitive solution in which all generators bid economically. Production cost at the competitive solution varies for FCP and aCHP.}
    \label{fig:P1_normproducercosts}
\end{figure}

\begin{figure}[tbh]
    \centering
    \includegraphics[width=.8\textwidth]{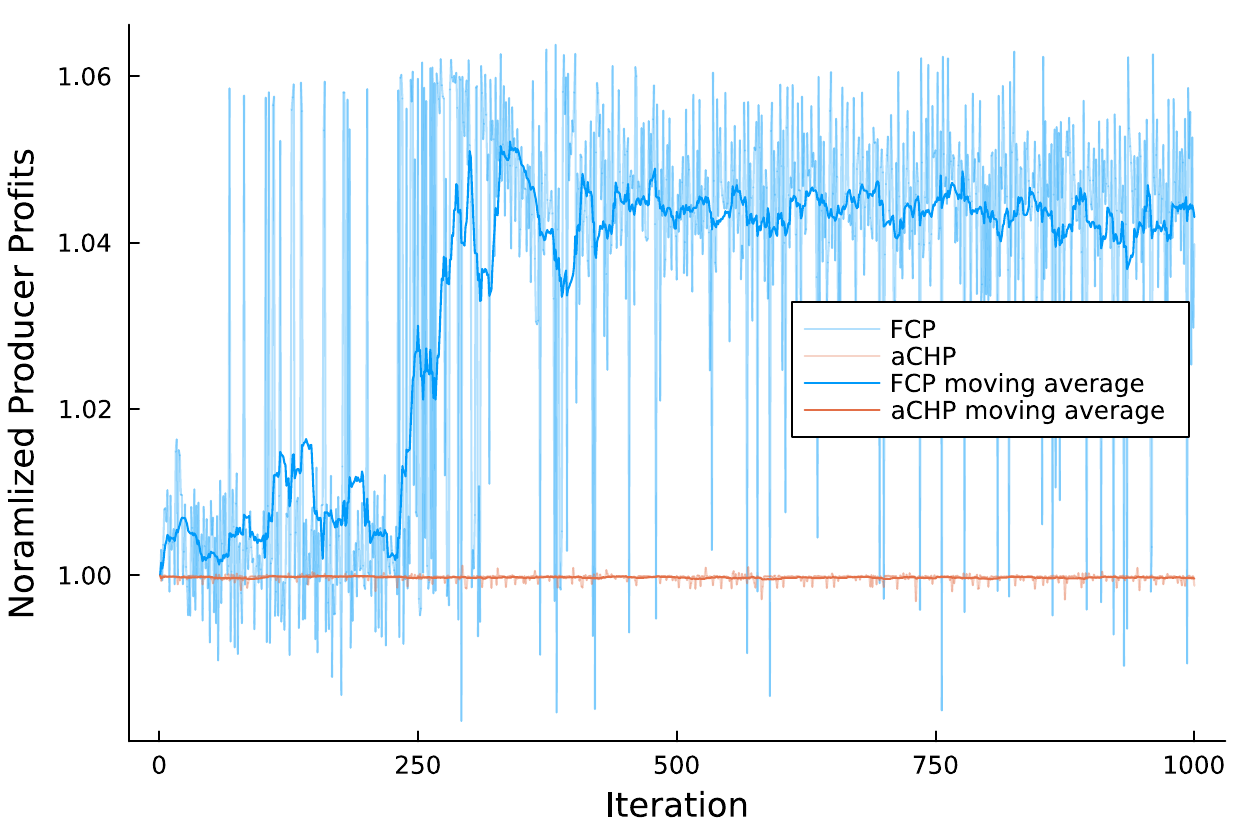}
    \caption{FERC1. Total producer profits normalized by profits at the competitive solution in which all generators bid economically for each pricing model.}
    \label{fig:P1_normproducerprofits}
\end{figure}

\begin{figure}[tbh]
    \centering
    \includegraphics[width=.8\textwidth]{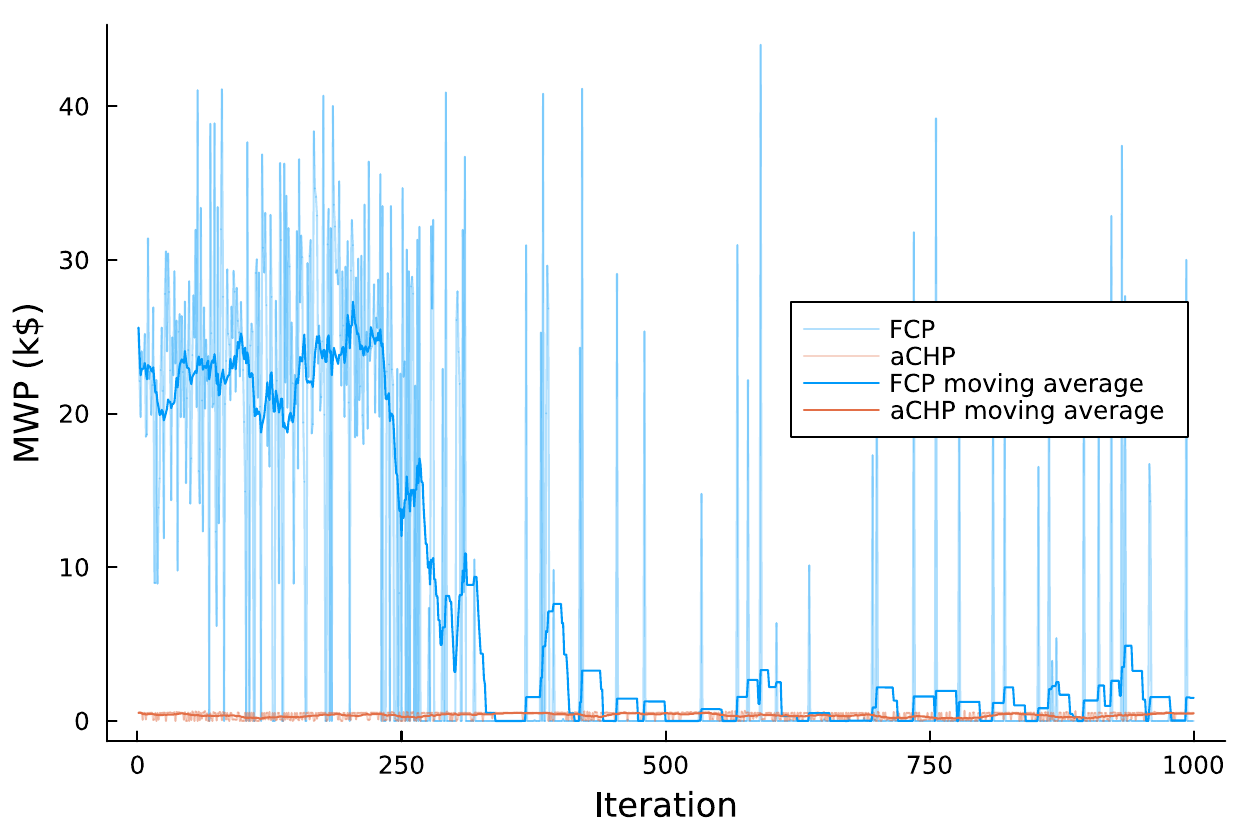}
    \caption{FERC1. Make-whole payments required for short-run cost recovery by pricing model.}
    \label{fig:P1_MWP}
\end{figure}

\begin{figure}[tbh]
    \centering
    \includegraphics[width=.8\textwidth]{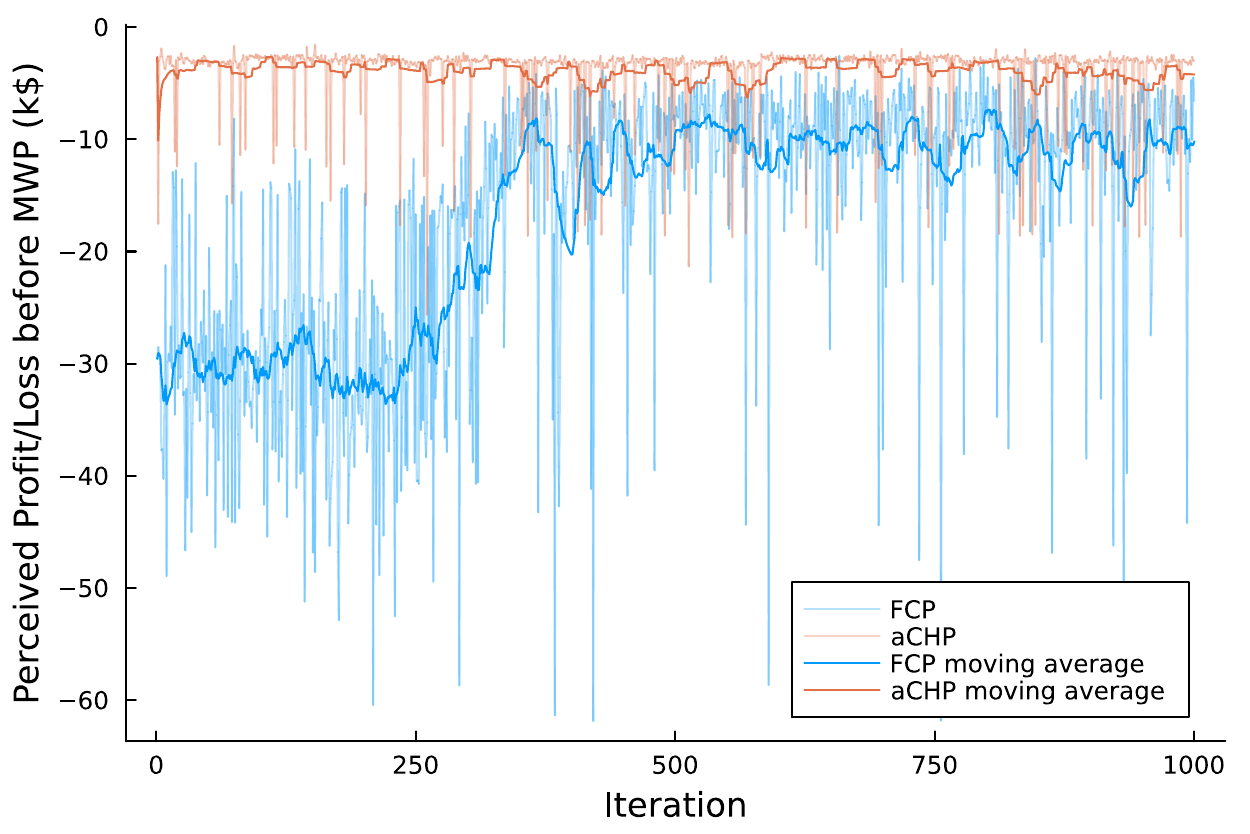}
    \caption{FERC1. Lost opportunity cost displayed as perceived profit or loss before MWP. Generators under FCP learn to bid strategically so as to lower LOC.}
    \label{fig:P1_LOC}
\end{figure}

\begin{figure}[tbh]
    \centering
    \includegraphics[width=.8\textwidth]{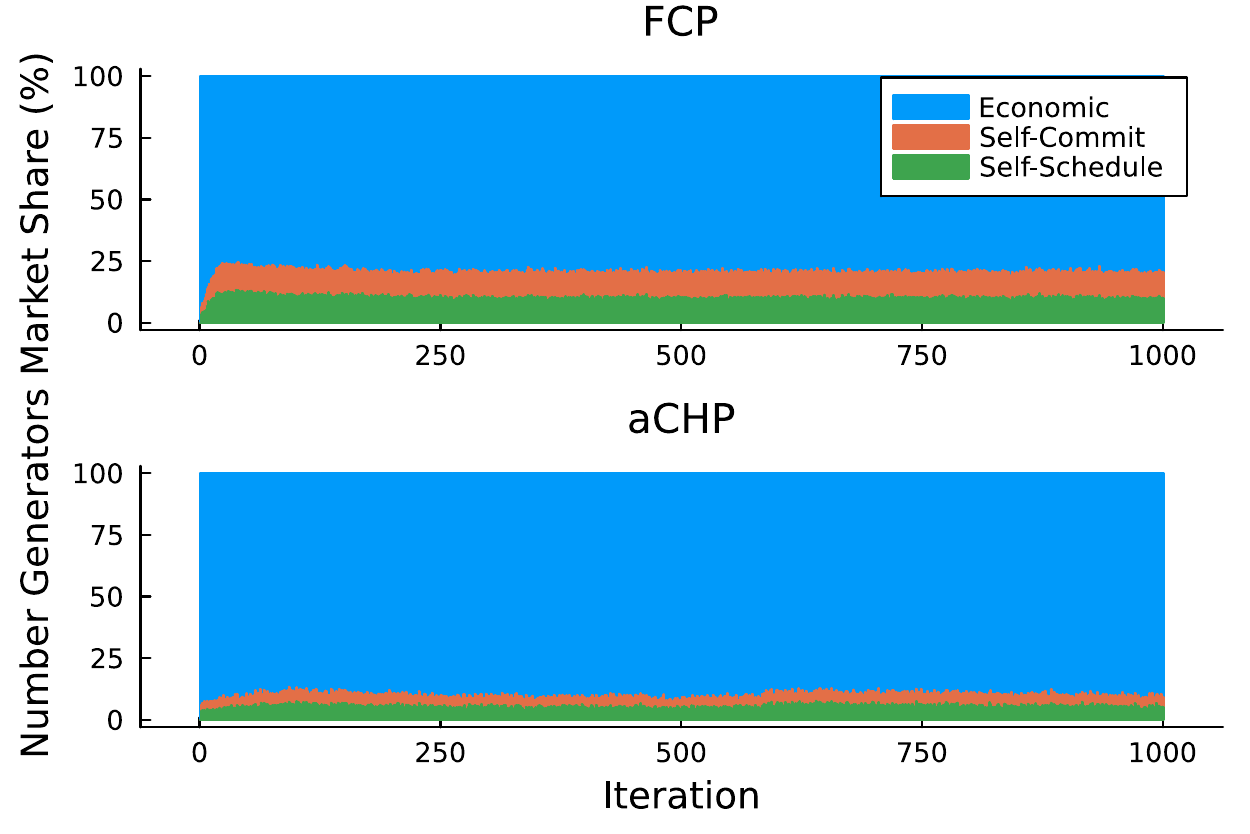}
    \caption{FERC1. Market share of each offer strategy per iteration as percentage of generators bidding each strategy.}
    \label{fig:P1_marketshareNumGens}
\end{figure}

\begin{figure}[tbh]
    \centering
    \includegraphics[width=.8\textwidth]{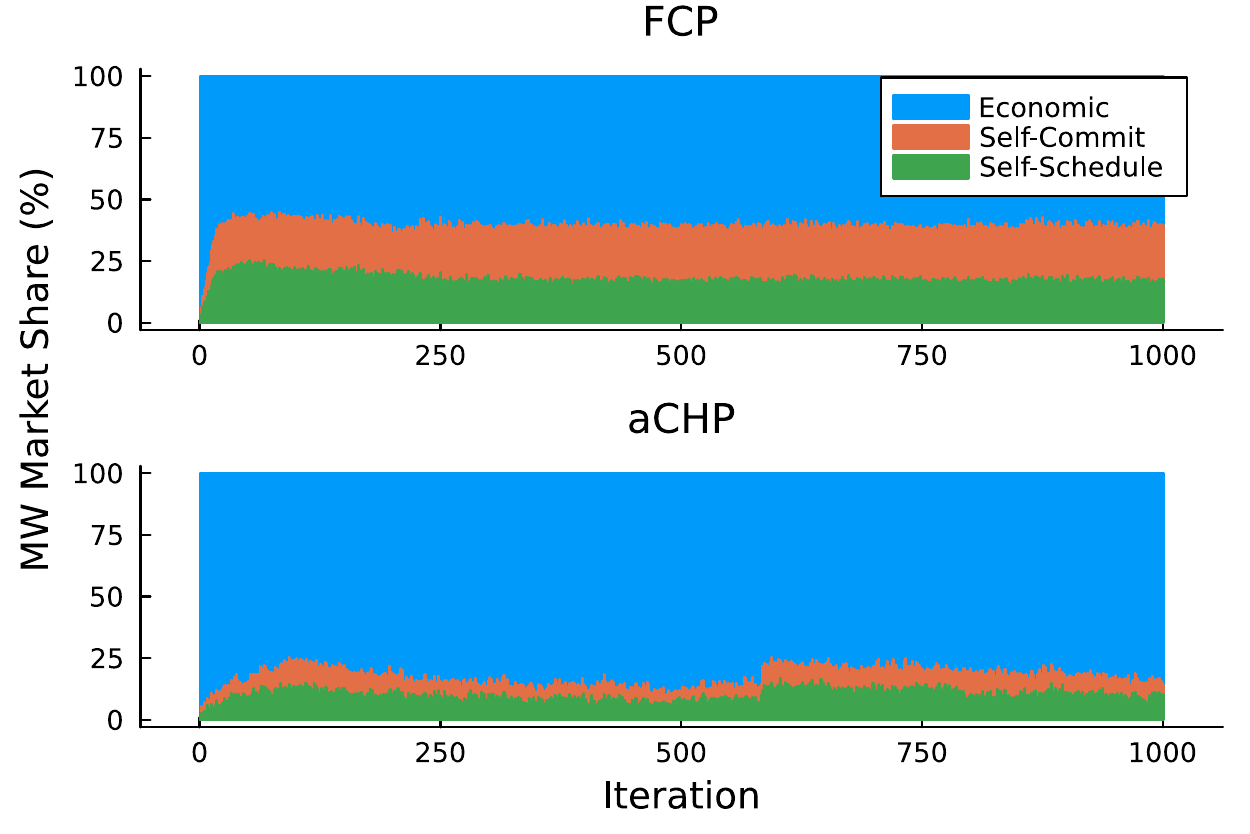}
    \caption{FERC1. Market share of each offer strategy per iteration as percentage of MW of total thermal generator capacity bidding each strategy.}
    \label{fig:P1_marketshareMW}
\end{figure}

\begin{figure}[tbh]
    \centering
    \includegraphics[width=.8\textwidth]{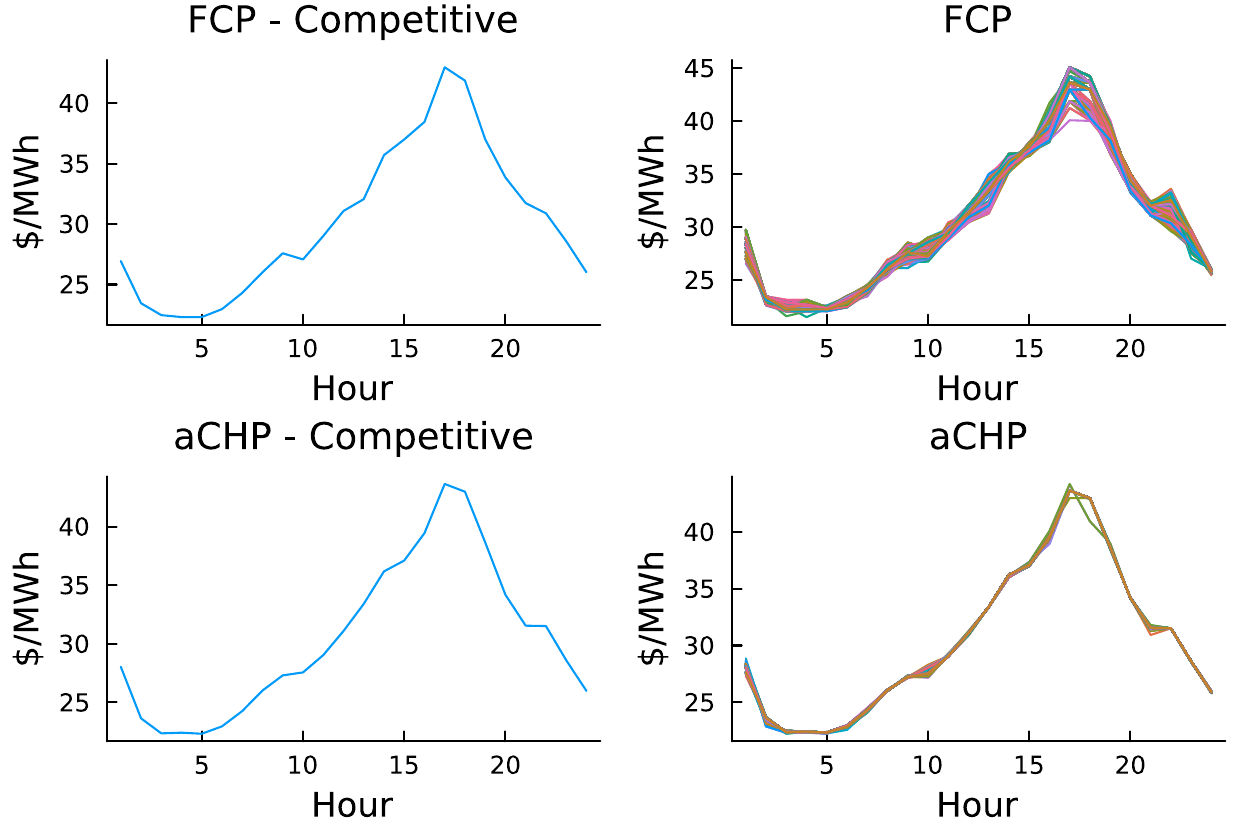}
    \caption{FERC1. Prices attained under the competitive solution in which all generators bid economically and prices attained over all iterations of strategic bidding.}
    \label{fig:P1_prices}
\end{figure}

\begin{figure}[tbh]
    \centering
    \includegraphics[width=.8\textwidth]{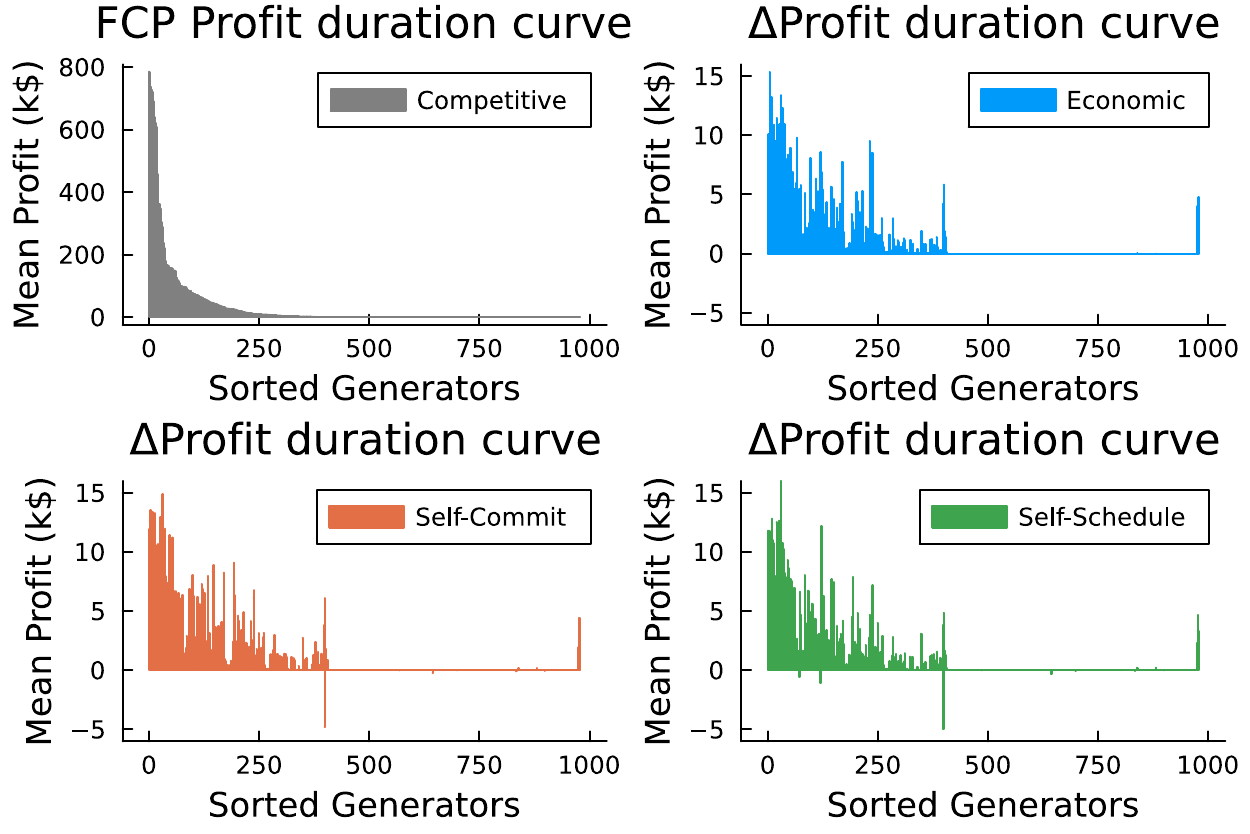}
    \caption{FERC1. Profit duration curve and deviations for FCP. Generators are sorted by profit achieved in the competitive outcome in which all generators submit economic bids. The difference between the mean profit achieved for each strategy in simulation and the profit achieved at the competitive solution is shown.}
    \label{fig:P1_profitdurationDeltaAllFCP}
\end{figure}

\begin{figure}[tbh]
    \centering
    \includegraphics[width=.8\textwidth]{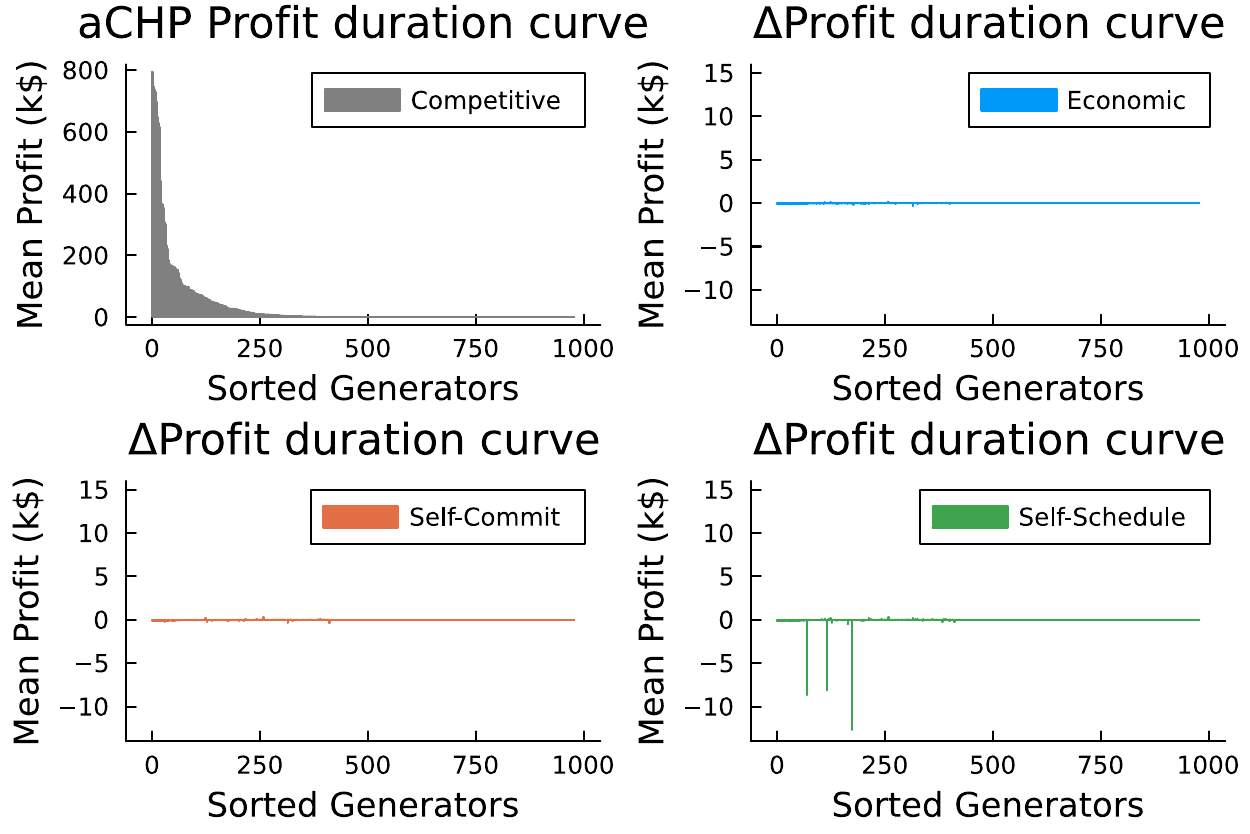}
    \caption{FERC1. Profit duration curve and deviations for aCHP. Generators are sorted by profit achieved in the competitive outcome in which all generators submit economic bids. The difference between the mean profit achieved for each strategy in simulation and the profit achieved at the competitive solution is shown.}
    \label{fig:P1_profitdurationDeltaAllaCHP}
\end{figure}

\begin{figure}[tbh]
    \centering
    \includegraphics[width=.8\textwidth]{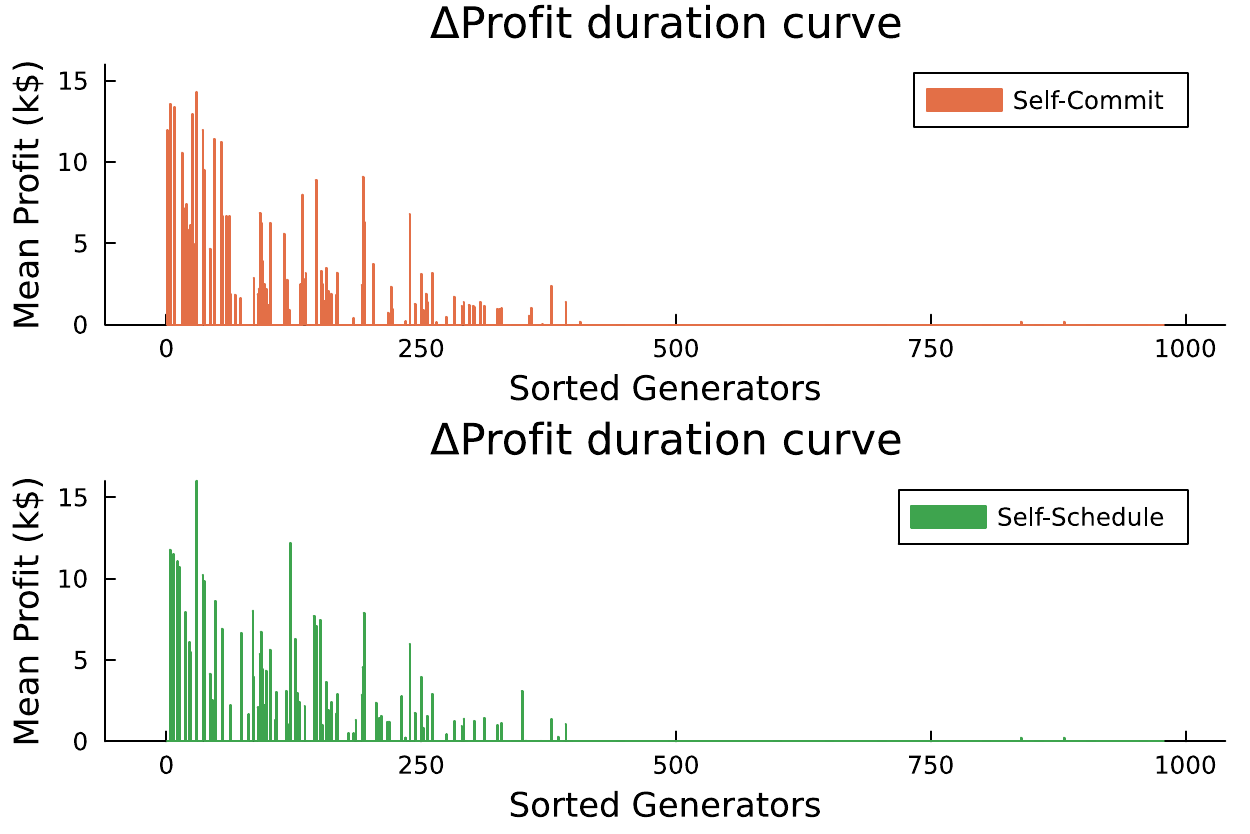}
    \caption{FERC1 (FCP). Deviations from the competitive solution profit duration curve for statistically significant adverse generators. Generators are sorted by profit achieved in the competitive outcome in which all generators submit economic bids. The difference between the mean profit achieved for each strategy in simulation and the profit achieved at the competitive solution is shown only for adverse generators.}
    \label{fig:P1_profitdurationDeltaAdverseFCP}
\end{figure}

\begin{figure}[tbh]
    \centering
    \includegraphics[width=.8\textwidth]{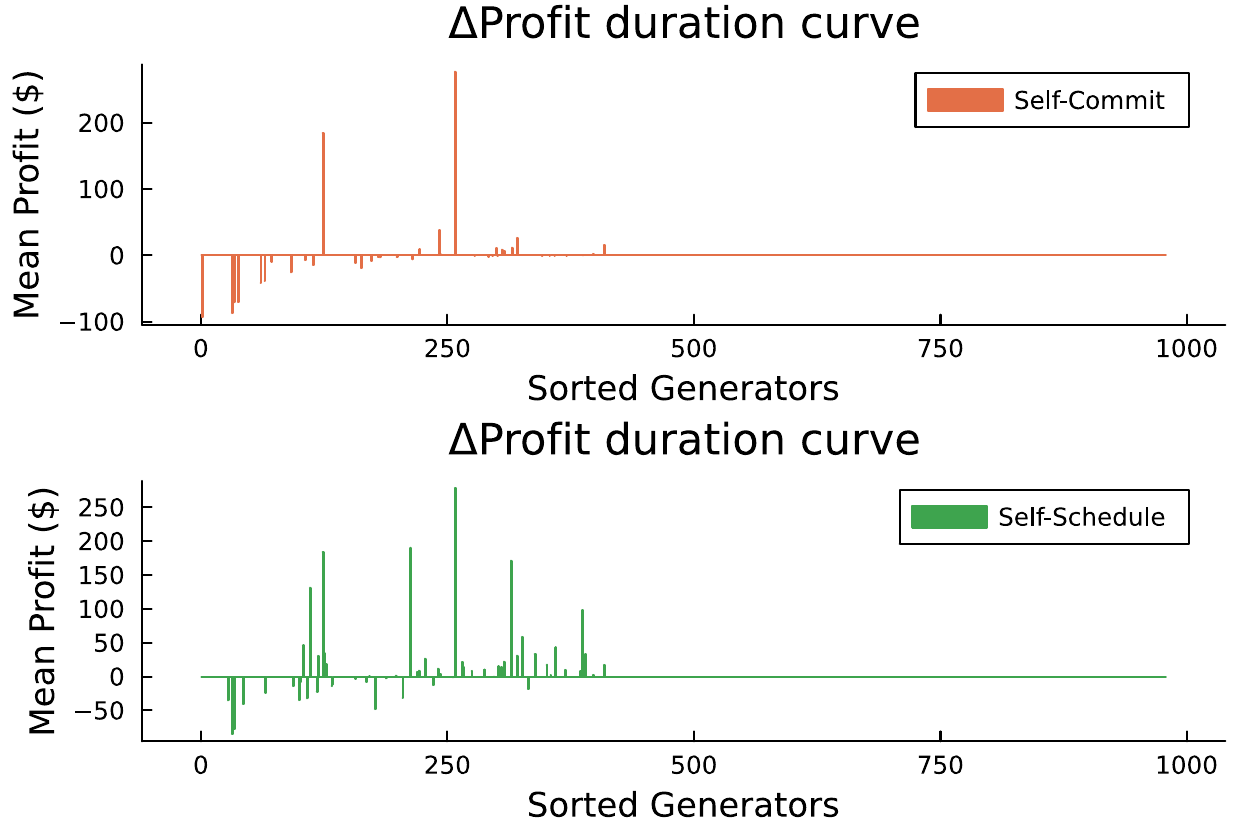}
    \caption{FERC1. (aCHP). Deviations from the competitive solution profit duration curve for statistically significant adverse generators. Generators are sorted by profit achieved in the competitive outcome in which all generators submit economic bids. The difference between the mean profit achieved for each strategy in simulation and the profit achieved at the competitive solution is shown only for adverse generators. Adverse generators are determined based on the expected payoff compared to economic bids in simulation, not payoff in the competitive solution, so $\Delta$ may be $<0$.}
    \label{fig:P1_profitdurationDeltaAdverseaCHP}
\end{figure}

\begin{figure}[tbh]
    \centering
    \includegraphics[width=.8\textwidth]{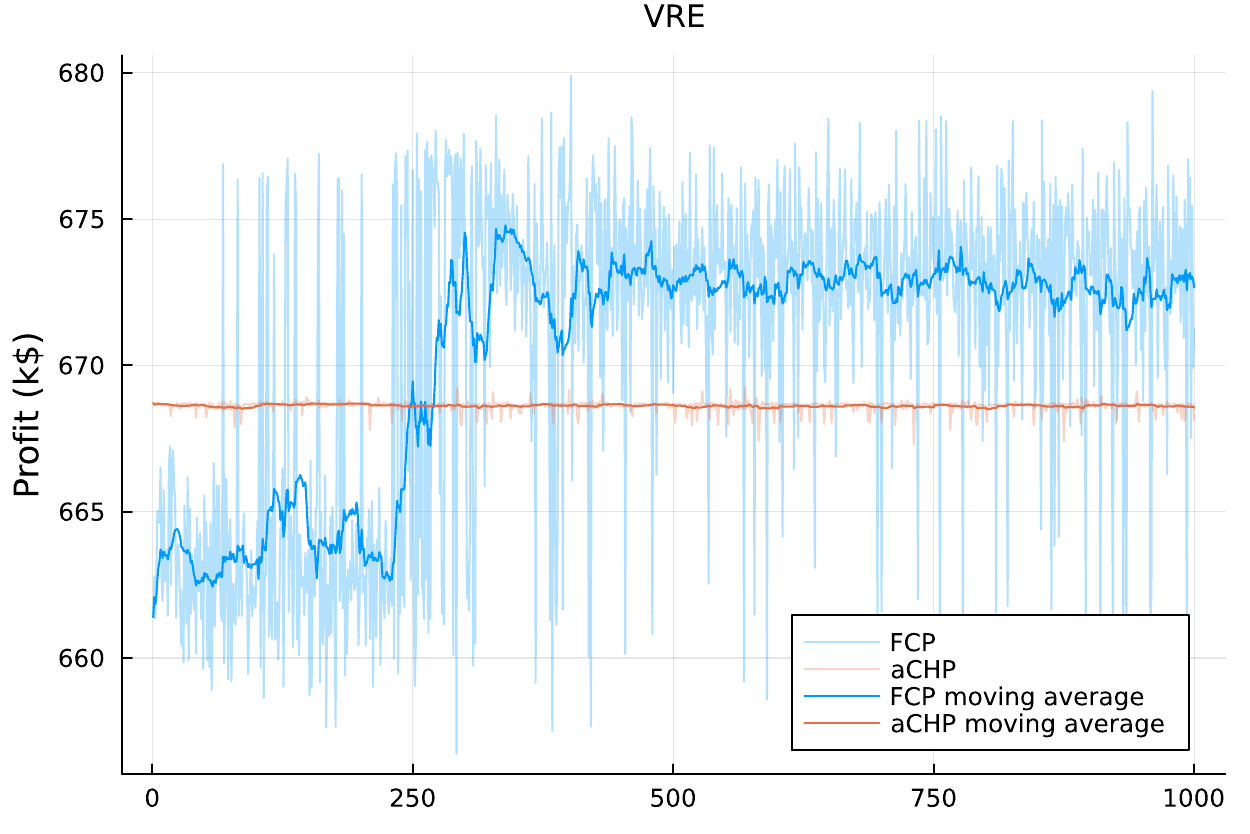}
    \caption{FERC1. Aggregate profit of VRE generators over iterations.}
    \label{fig:P1_profitVRE}
\end{figure}

Results for FERC1 demonstrate that generators are able to learn to bid strategically in a way that increases producer profits and the cost to consumers. Figure \ref{fig:P1_consumercosts} shows that while the cost to consumers changes negligibly with aCHP when generators can self-commit or self-schedule, the cost to consumers under FCP rises. At the competitive solution, the cost to consumers with FCP is 1.2\% lower than with aCHP. However, the mean cost to consumers for FCP in the final 500 iterations is 0.76\% higher than the mean cost to consumers for aCHP in the final 500 iterations. The average cost to consumers under FCP settles at approximately 2\% higher with strategic bidding than the competitive solution, as shown in Figure \ref{fig:P1_normconsumercosts}. Under FCP, total producer profits increase on average in the final 500 iterations by 4.4\% compared to the competitive solution, as shown in Figure \ref{fig:P1_normproducerprofits}. The efficiency of the system operator's solution with strategic bidding can decrease with strategic bids, but only very slightly, as shown in Figure \ref{fig:P1_normproducercosts}. Note that each iteration only solves to a MIP gap of 0.001\%, so some iterations can also achieve slightly lower producer costs than the competitive solution in the first iteration. 

Figures \ref{fig:P1_marketshareNumGens} and \ref{fig:P1_marketshareMW} show the market share of each strategy per iteration in terms of number of units and total MW bidding each offer strategy. For FCP, generators that self-commit or self-schedule represent 38.8\% of the total thermal generator capacity on average in the final 500 iterations. For aCHP, 18.3\% of available MW self-commit or self-schedule on average in the final 500 iterations.

However, not all generators that bid strategically profit because of their strategy; some randomly learn to bid strategically due to the behavior of other generators that influence the price. Tables \ref{tab:FERCAdverseStratBidsFCP} and \ref{tab:FERCAdverseStratBidsaCHP} show the number of statistically significant adverse strategic bidders for each pricing model, and Tables \ref{tab:FERCAdverseStratBidsMWFCP} and \ref{tab:FERCAdverseStratBidsMWaCHP} show the total MW represented by these adverse strategic bidders. Under FCP, generators representing 24.7\% of thermal capacity profit by either self-scheduling or self-committing. Under aCHP, this number is 13.9\%. Total excess profits for adverse bidders are given in Table \ref{tab:FERCAdverseStratBidsPayoffs}. Excess profits are defined in Section \ref{sec:stylizedtestcase} as the difference for strategic generators between the mean profit for the strategic bidding strategy with the highest payoff (either $\overline{X}_{selfcomm}$ or $\overline{X}_{selfsched}$) and $\overline{X}_{eco}$, the mean profit when bidding economically. Total excess profits for FCP are 0.73\% of FCP competitive profits, while total excess profits for aCHP are only 0.01\% of aCHP competitive profits. 

Figure \ref{fig:P1_MWP} shows the total MWP required for cost recovery in each iteration. The aCHP model has higher prices in the competitive solution in the first iteration, and thus has lower MWP requirements. Since generators in FCP learn to strategically bid to increase the price, MWP fall over iterations. When generators are learning to increase their profits by bidding strategically, they are learning to decrease LOC. Figure \ref{fig:P1_LOC} shows that while LOC has no trend for aCHP, LOC under FCP decreases over the iterations.

The prices found with strategic bidding vary far more under FCP than under aCHP, leading to increased profit potential. Competitive prices and the range of prices found via strategic bidding for each pricing model are shown in Figure \ref{fig:P1_prices}.

The profit duration curve under competitive conditions and deviations from this curve for each generator under each bidding strategy are shown in Figures \ref{fig:P1_profitdurationDeltaAllFCP} and \ref{fig:P1_profitdurationDeltaAllaCHP}. For FCP, many generators benefit from the allowance of strategic bidding and resulting higher prices even if they cannot themselves impact the price. Typically it is generators that were already highly profitable under competitive conditions that have the highest payoffs under strategic conditions. For aCHP, there is no additional profit potential relative to competitive profits, and some potential for losses, indicating that aCHP incentives generators not to self-schedule if they could impact the price. Figures \ref{fig:P1_profitdurationDeltaAdverseFCP} and \ref{fig:P1_profitdurationDeltaAdverseaCHP} show the deviation from competitive profits for statistically significant adverse bidders. Under FCP, again the generators with the highest competitive profits benefit the most from bidding strategically. Under aCHP, there is little to no increased payoff relative to competitive conditions (note the different axis scale in Figure \ref{fig:P1_profitdurationDeltaAdverseaCHP}). Because adverse generators are determined based on the expected payoff of a strategic bid compared to economic bids while other generators are also bidding strategically instead of the payoff in competitive conditions, some of the deviations from competitive profits are negative.

The zero-marginal cost VRE generator also benefits from the higher profits induced by strategic thermal generators under FCP. Figure \ref{fig:P1_profitVRE} shows the total profit achieved under CHP changes negligibly, but the total profit achieved under FCP grows as the thermal generators determine optimal strategic bidding strategies.

 \begin{table}[H]
    \centering
    \caption{Adverse Strategic Bids (FCP)}
    \begin{tabular}{ cccc } 
     \toprule
     &\multicolumn{3}{c}{Number of Generators}\\
       &\multicolumn{3}{c}{$\overline{X}_{eco} < (\overline{X}_{selfsched}$ OR $ \overline{X}_{selfcomm}) $} and $p<0.05$\\
      Strategy & FERC1 & FERC2& FERC3\\
     \midrule
      Self-Commit & 90 (9.2\%)  & 61  (6.2\%) & 94 (10.1\%)\\
      Self-Schedule & 82 (8.4\%) & 66 (6.7\%) & 103 (11.0\%)\\
      Total (Unique) & 119 (12.2\%) & 93 (9.5\%) & 136 (14.6\%)\\
    \bottomrule
    \end{tabular}
    \label{tab:FERCAdverseStratBidsFCP}

    \vspace{.5cm}
    \caption{Adverse Strategic Bids (aCHP)}
    \begin{tabular}{ cccc } 
     \toprule
     &\multicolumn{3}{c}{Number of Generators}\\
       &\multicolumn{3}{c}{$\overline{X}_{eco} < (\overline{X}_{selfsched}$ OR $ \overline{X}_{selfcomm}) $} and $p<0.05$\\
      Strategy & FERC1 & FERC2& FERC3\\
     \midrule
      Self-Commit & 39 (4.0\%) & 19 (1.9\%) & 16 (1.7\%)\\
      Self-Schedule & 63 (6.4\%)  & 37 (3.8\%) & 26 (2.8\%)\\
      Total (Unique) & 90 (9.2\%) &  47 (4.8\%)& 35 (3.7\%)\\
    \bottomrule
    \end{tabular}
    \label{tab:FERCAdverseStratBidsaCHP}
\end{table}

 \begin{table}[H]
    \centering
    \caption{Adverse Strategic Bids (FCP)}
    \begin{tabular}{ cccc } 
     \toprule
     &\multicolumn{3}{c}{Generator Capacity (GW)}\\
       &\multicolumn{3}{c}{$\overline{X}_{eco} < (\overline{X}_{selfsched}$ OR $ \overline{X}_{selfcomm}) $} and $p<0.05$\\
      Strategy & FERC1 & FERC2& FERC3\\
     \midrule
      Self-Commit & 32.8  (18.5\%)  & 28.6 (16.1\%) & 50.7 (28.0\%)\\
      Self-Schedule & 30.2 (17.0\%) & 29.4 (16.6\%) & 50.4 (27.9\%)\\
      Total (Unique) & 43.9  (24.7\%) &42.4 (23.9\%) & 67.3 (37.2\%)\\
    \bottomrule
    \end{tabular}
    \label{tab:FERCAdverseStratBidsMWFCP}

    \vspace{.5cm}
    \caption{Adverse Strategic Bids (aCHP)}
    \begin{tabular}{ cccc } 
     \toprule
     &\multicolumn{3}{c}{Generator Capacity MW }\\
       &\multicolumn{3}{c}{$\overline{X}_{eco} < (\overline{X}_{selfsched}$ OR $ \overline{X}_{selfcomm}) $} and $p<0.05$\\
      Strategy & FERC1 & FERC2& FERC3\\
     \midrule
      Self-Commit & 9.2 (5.2\%)  & 6.1 (3.4\%) & 3.4 (1.9\%)\\
      Self-Schedule & 19.0  (10.7\%) & 12.4 (7.0\%) & 7.3 (4.0\%)\\
      Total (Unique) & 24.7  (13.9\%) & 14.0 (7.9\%) &  9.8 (5.4\%)\\
    \bottomrule
    \end{tabular}
    \label{tab:FERCAdverseStratBidsMWaCHP}
\end{table}

\begin{table}[H]
    \centering
    \caption{Adverse Strategic Bids Payoffs}
    \begin{tabular}{ cccc } 
     \toprule
       &\multicolumn{3}{c}{ Total excess profit (\% competitive profits)}\\
      Pricing Model& FERC1 & FERC2& FERC3\\
     \midrule
     FCP & \$236.0k (0.73\%) & \$32.7k (0.09\%) & \$204.2k (0.60\%)\\ 
     aCHP & \$2.6k (0.01\%) & \$1.0k (0.00\%) & \$1.4k (0.00\%)\\ 
    \bottomrule
    \end{tabular}
    \label{tab:FERCAdverseStratBidsPayoffs}
\end{table}


\begin{figure}[tbh]
    \centering
    \includegraphics[width=.8\textwidth]{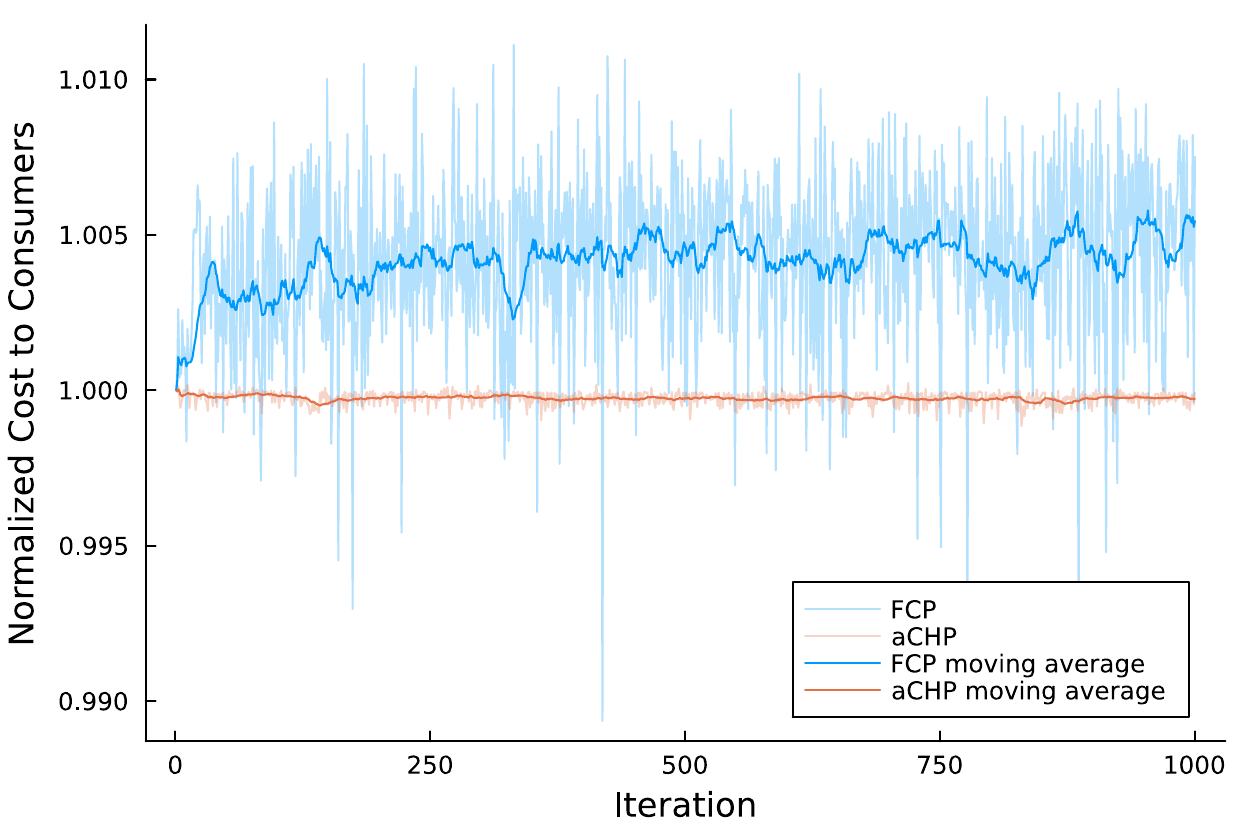}
    \caption{FERC2. Cost to consumers over iterations normalized by cost to consumers at the competitive solution in which all generators bid economically for each pricing model.}
    \label{fig:P2_normconsumercosts}
\end{figure}

\begin{figure}[tbh]
    \centering
    \includegraphics[width=.8\textwidth]{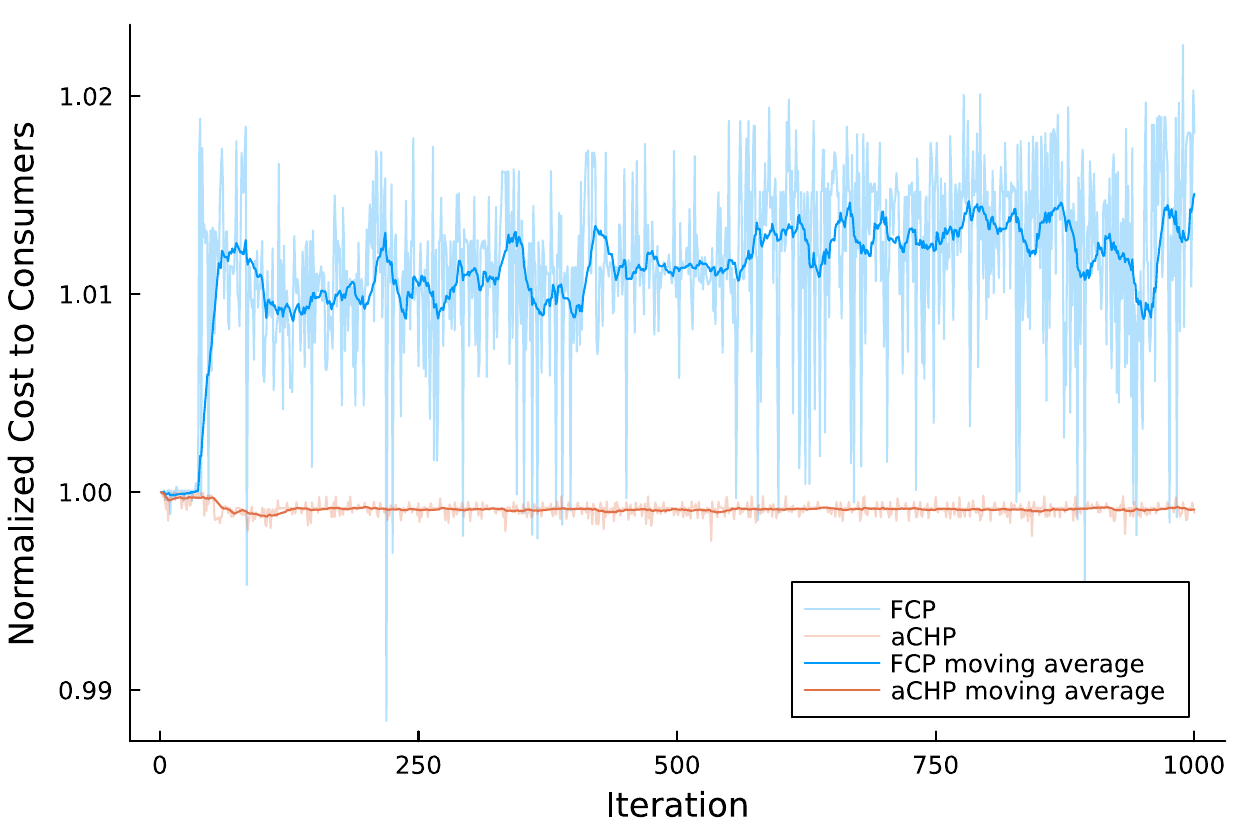}
    \caption{FERC3. Cost to consumers over iterations normalized by cost to consumers at the competitive solution in which all generators bid economically for each pricing model.}
    \label{fig:P3_normconsumercosts}
\end{figure}

\begin{figure}[tbh]
    \centering
    \includegraphics[width=.8\textwidth]{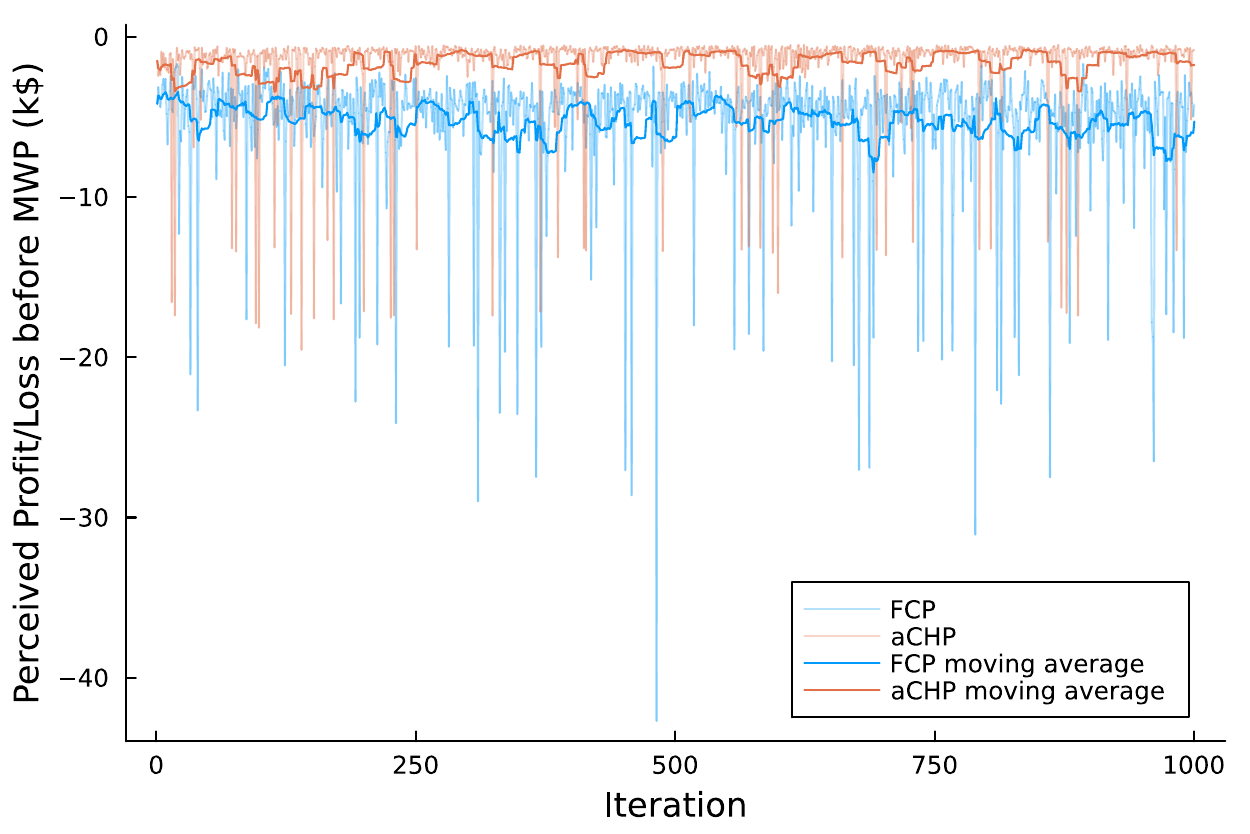}
    \caption{FERC2. Lost opportunity cost displayed as perceived profit or loss before MWP.}
    \label{fig:P2_LOC}
\end{figure}

\begin{figure}[tbh]
    \centering
    \includegraphics[width=.8\textwidth]{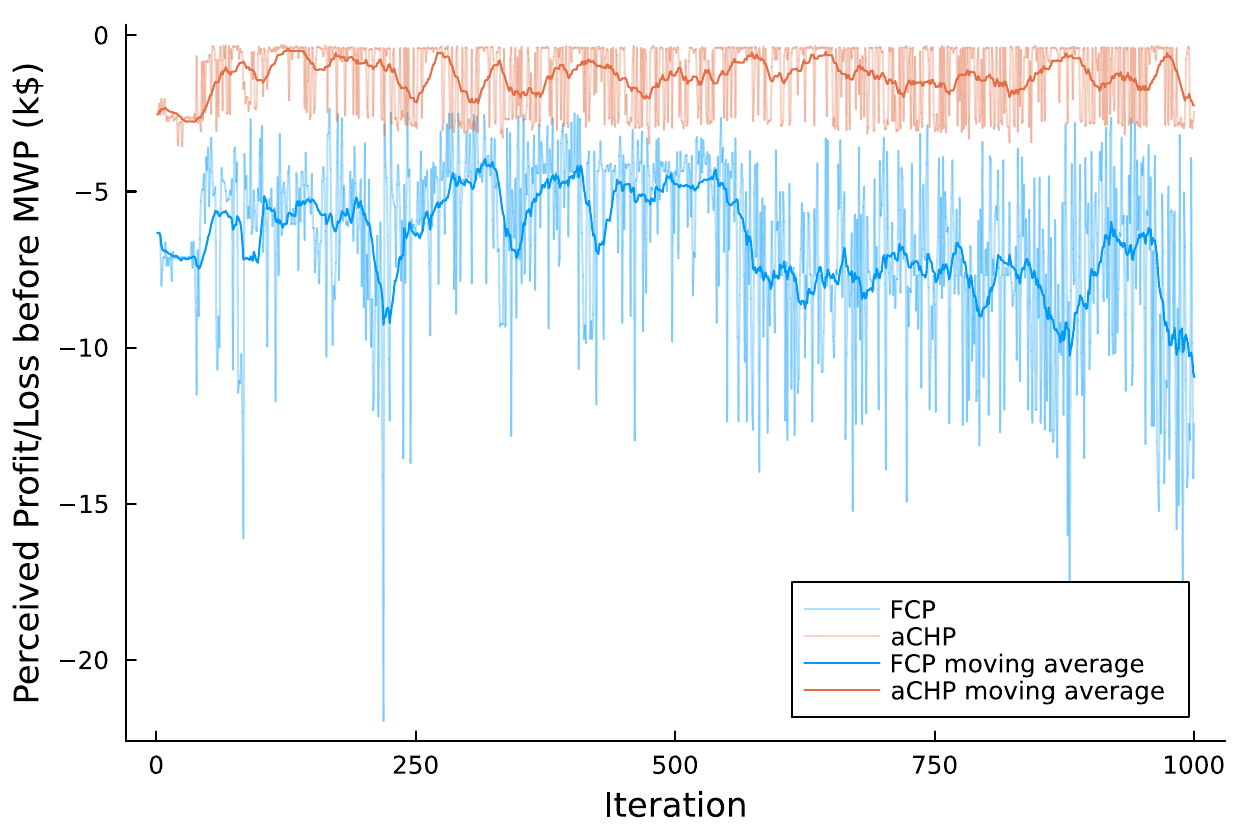}
    \caption{FERC3. Lost opportunity cost displayed as perceived profit or loss before MWP.}
    \label{fig:P3_LOC}
\end{figure}

\begin{figure}[tbh]
    \centering
    \includegraphics[width=.8\textwidth]{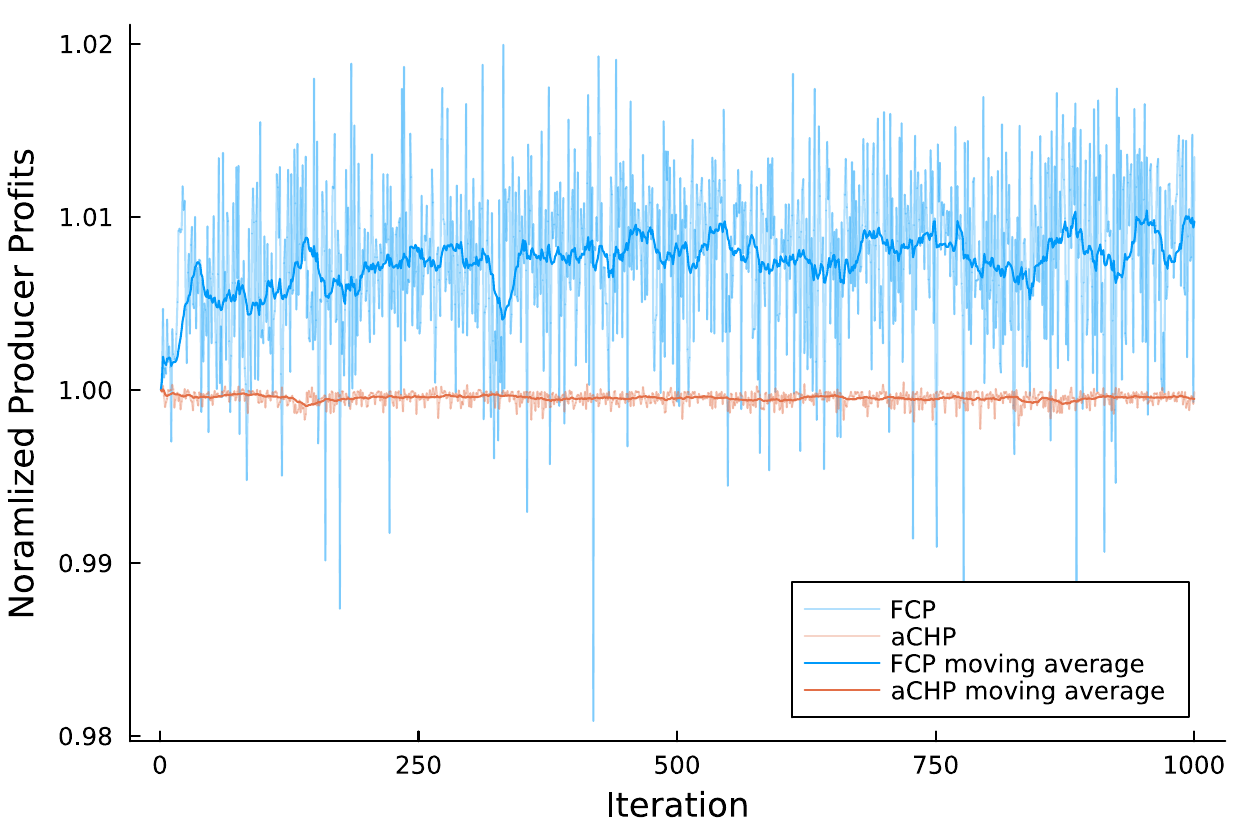}
    \caption{FERC2. Total producer profits normalized by profits at the competitive solution in which all generators bid economically for each pricing model.}
    \label{fig:P2_normproducerprofits}
\end{figure}

\begin{figure}[tbh]
    \centering
    \includegraphics[width=.8\textwidth]{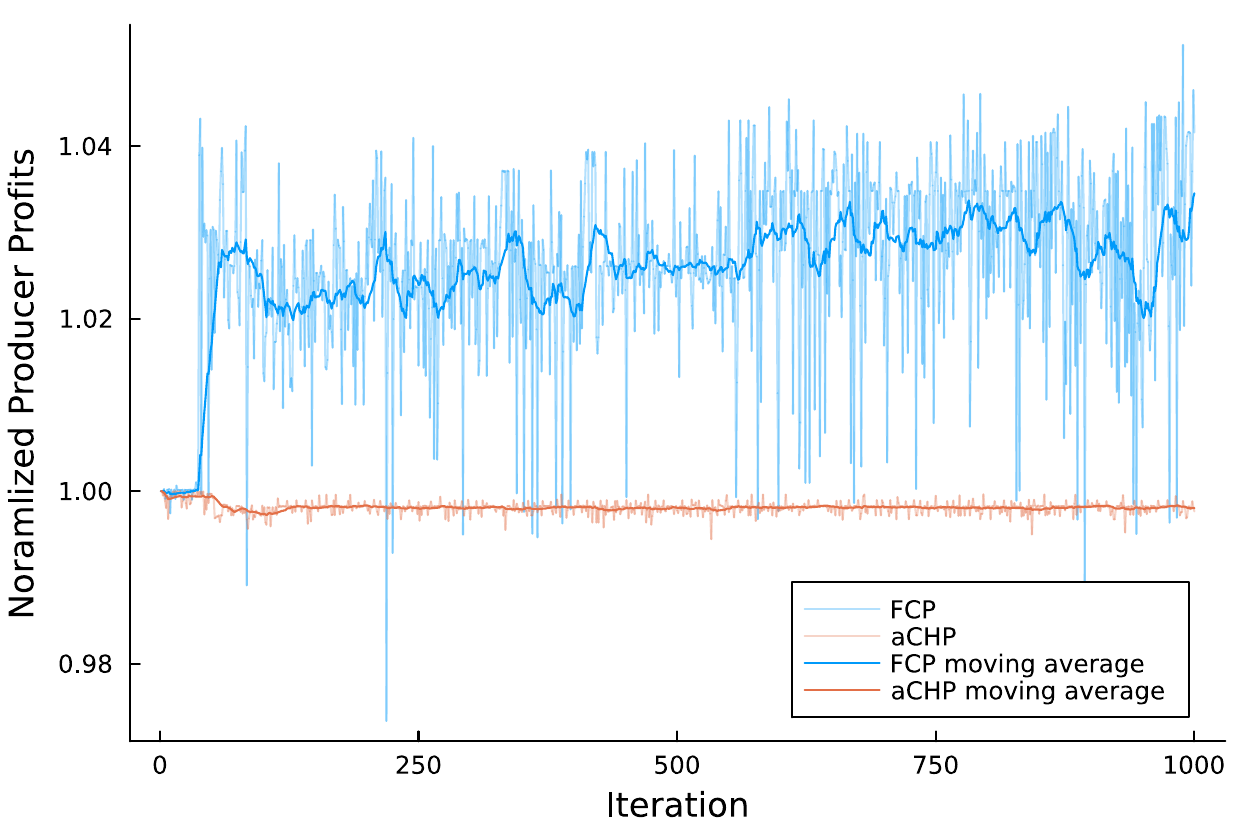}
    \caption{FERC3. Total producer profits normalized by profits at the competitive solution in which all generators bid economically for each pricing model.}
    \label{fig:P3_normproducerprofits}
\end{figure}

\begin{figure}[tbh]
    \centering
    \includegraphics[width=.8\textwidth]{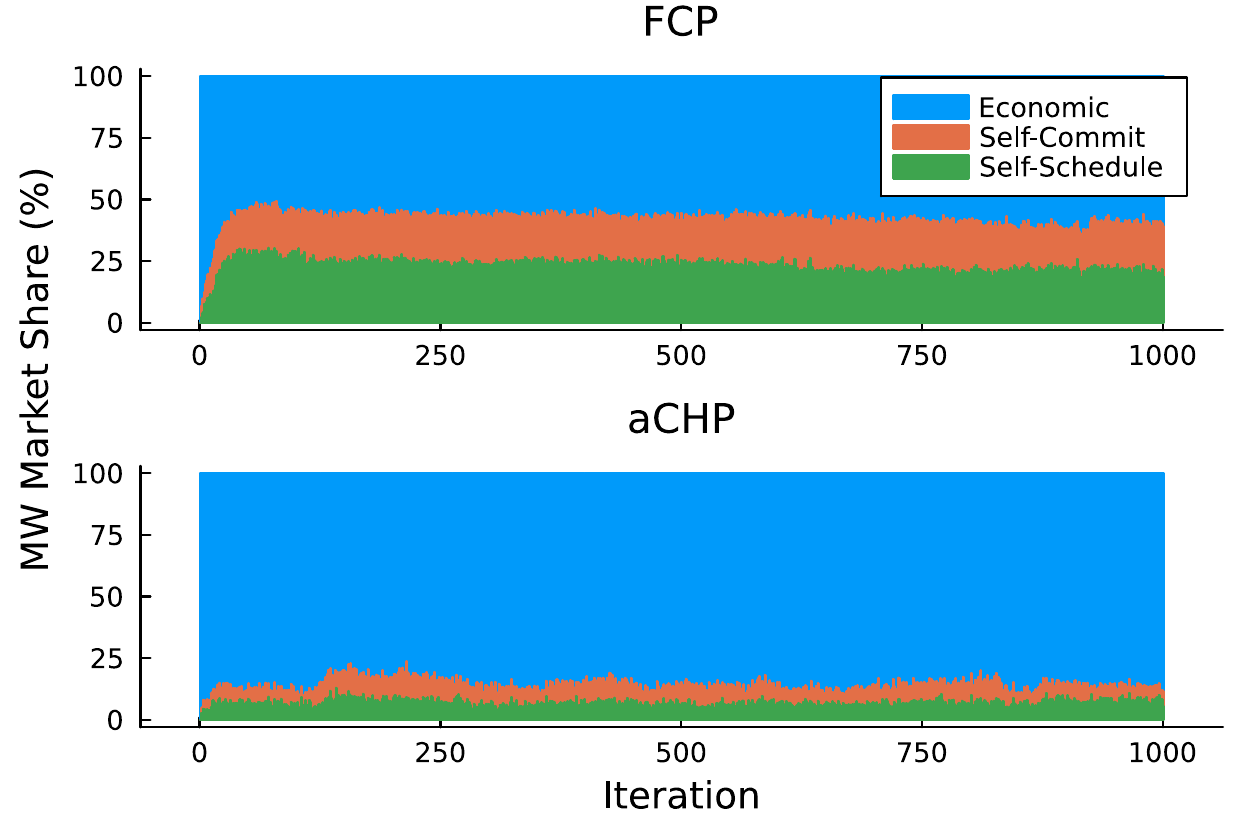}
    \caption{FERC2. Market share of each offer strategy per iteration as percentage of MW of total thermal generator capacity bidding each strategy.}
    \label{fig:P2_marketshareMW}
\end{figure}

\begin{figure}[tbh]
    \centering
    \includegraphics[width=.8\textwidth]{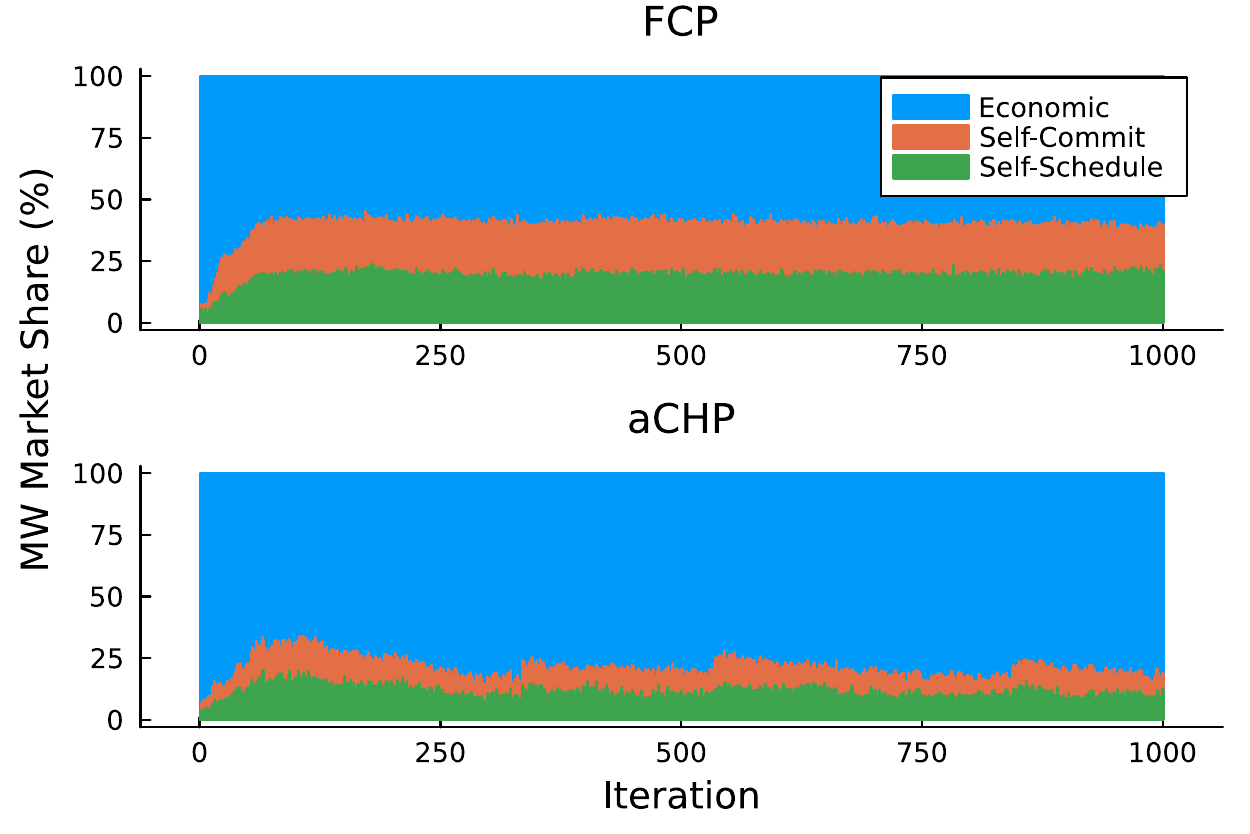}
    \caption{FERC3. Market share of each offer strategy per iteration as percentage of MW of total thermal generator capacity bidding each strategy.}
    \label{fig:P3_marketshareMW}
\end{figure}

\begin{figure}[tbh]
    \centering
    \includegraphics[width=.8\textwidth]{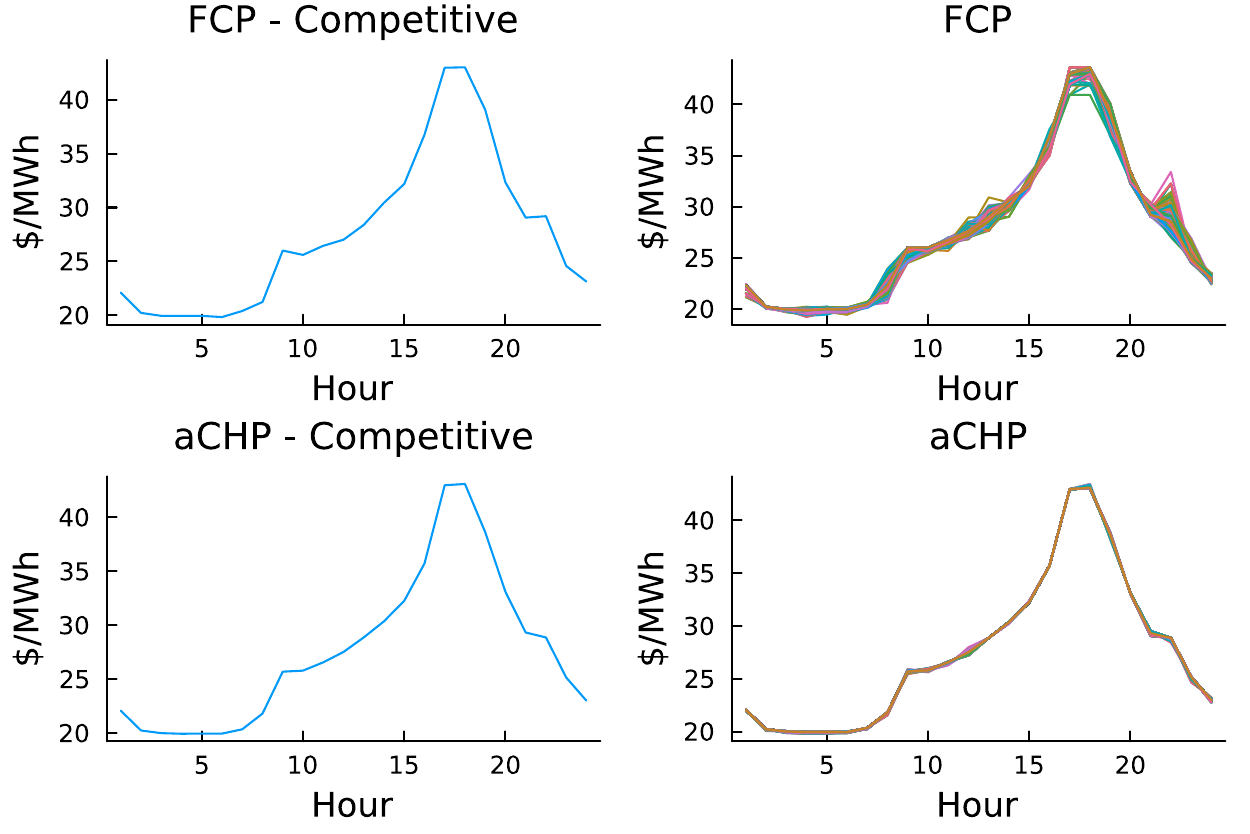}
    \caption{FERC2. Prices attained under the competitive solution in which all generators bid economically and prices attained over all iterations of strategic bidding.}
    \label{fig:P2_prices}
\end{figure}

\begin{figure}[tbh]
    \centering
    \includegraphics[width=.8\textwidth]{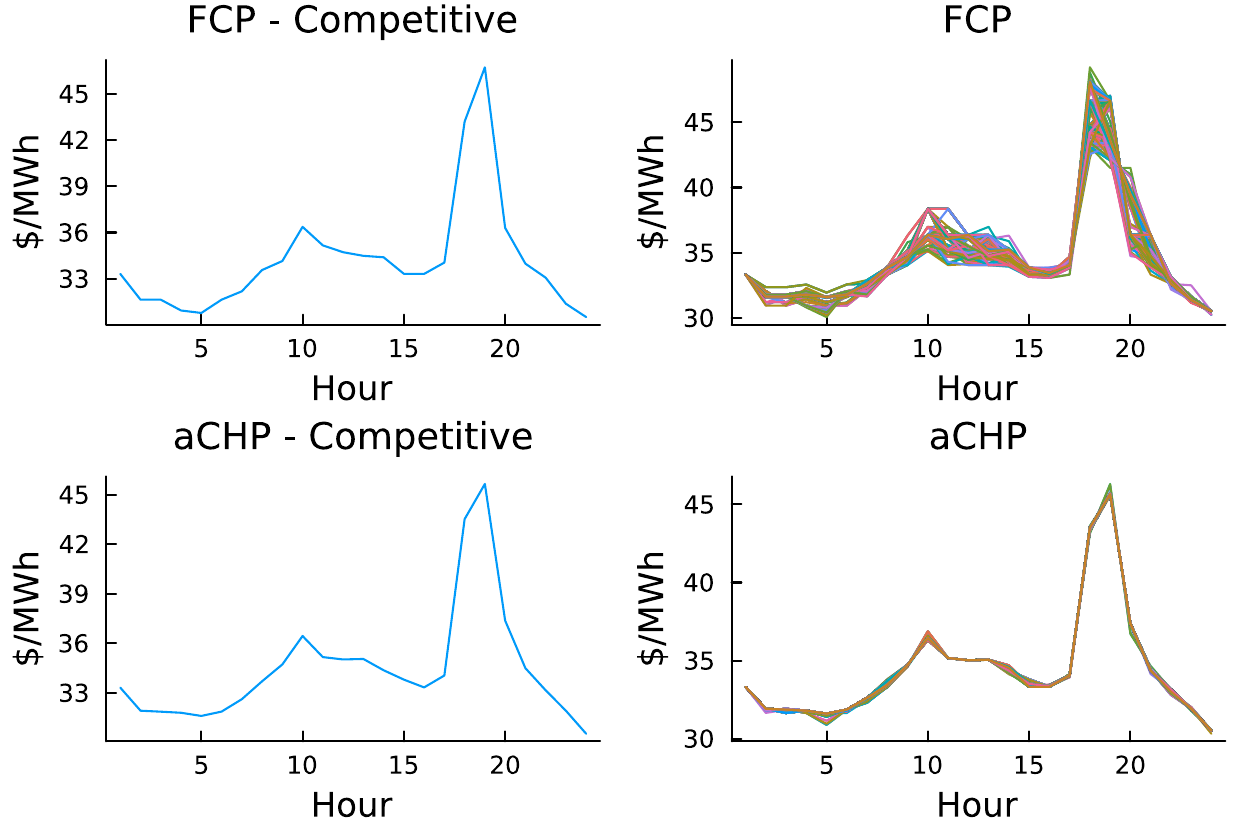}
    \caption{FERC3. Prices attained under the competitive solution in which all generators bid economically and prices attained over all iterations of strategic bidding.}
    \label{fig:P3_prices}
\end{figure}


FERC2 is the same case as FERC1 except a large quantity of wind was added. This means the resource mix is far from the long-run adapted resource mix, so we would expect to see less opportunity for profits for thermal generators, and that less thermal generators overall will be committed. FERC3 is another low wind case, but on a winter instead of summer day.

Tables \ref{tab:FERCAdverseStratBidsFCP} and \ref{tab:FERCAdverseStratBidsaCHP} show the number of statistically significant adverse strategic bidders for each pricing model, Tables \ref{tab:FERCAdverseStratBidsMWFCP} and \ref{tab:FERCAdverseStratBidsMWaCHP} show the total MW represented by these adverse strategic bidders, and Table \ref{tab:FERCAdverseStratBidsPayoffs} shows the total excess profit. Under FCP, FERC2 with has a similar amount of MW that are statistically significant adverse bidders as FERC1 (23.9\% vs 24.7\%), but the total excess profits are lower (0.09\% vs 0.73\%). FERC3 has a higher share of MW as adverse bidders at 37.2\%, and total excess profits of 0.60\%. The MW of adverse bidders for aCHP is lower than FCP in both cases, and the excess profits for both is 0.00\%, i.e., the payoffs from strategic bidding, while statistically significant, are very small.

While the normalized cost to consumers under FCP in FERC2 only increases by approximately 0.05\% (Figure \ref{fig:P2_normconsumercosts}), the normalized cost to consumers in FERC3 increases by approximately 1.5\% (Figure \ref{fig:P3_normconsumercosts}). The increase in normalized producer profits for FERC2 shown in Figure \ref{fig:P2_normproducerprofits} is also lower than in FERC1 (less than 1\% vs 4.4\%), while the average increase in FERC3 in the final 500 iterations is 2.9\%, shown in Figure \ref{fig:P3_normproducerprofits}.

The LOC in the competitive solution for FCP is far lower in FERC2 and FERC3 compared to FERC1, and there is no trend of decreasing LOC with learning as with FERC1. Figures \ref{fig:P2_LOC} and \ref{fig:P3_LOC} show LOC for FERC2 and FERC3. The market share of each offer strategy for FERC2 and FERC3, shown in Figures 
 \ref{fig:P2_marketshareMW} and \ref{fig:P3_marketshareMW}, are similar to those found in FERC1. The range of prices found under FCP with strategic bidding varies more and reaches higher values in FERC3 than FERC2, as shown in Figures \ref{fig:P2_prices} and \ref{fig:P3_prices}. The peak price in particular varies more for FERC1 and FERC3 than FERC2, with significant added wind.

\section{Conclusion}

In a market with non-convex costs, market power can be exercised by self-committing/scheduling. In electricity markets that permit self-commitment and self-scheduling, generators can learn to bid strategically to increase their profits using reinforcement learning without knowledge of the costs or strategies of other generators. While the FCP pricing model provides incentives to adversely self-commit or self-schedule, convex hull pricing provides minimal incentives to deviate from the socially-optimal dispatch solution.

 Using a realistic test system, we find that when LOC is high under FCP, generators can learn to bid strategically to increase their profits and lower LOC. In one test case strategic bidding decreased total system LOC under FCP by approximately 2/3. Generators who are able to adversely self-commit or self-schedule to increase their profits tend to be generators who were already highly profitable under competitive conditions. However, many generators benefit from the higher prices induced by strategic generators. In our simulations, approximately 40\% of thermal generating capacity under FCP learned to self-schedule or self-commit, similar to the levels in markets today. Importantly, we are finding this behavior without the presence of long lead time scheduling constraints or take-or-pay fuel contracts that are typically used to explain this behavior. Of this amount, between 24\%-38\% across cases increased their payoff by bidding strategically rather than bidding economically in a statistically significant manner while other generators were bidding strategically. Cost to consumers under aCHP is higher in competitive conditions, but cost to consumers under FCP is higher in strategic conditions. Producer profits increased in cases with low wind 2.9\% and 4.4\%, while they increased in a case with significant added wind (and thus less profit potential for the same resource mix of thermal generators) only 1\%.

The ability of generators to adversely self-commit or self-schedule depends on the number and characteristics of the generators in the system. If there is significant excess thermal generation capacity, the profit potential is lower. More work is needed to explore how different resource mixes and net load profiles influence the potential for adverse self-commitments and self-scheduling.

Future work should consider how transmission constraints and reserve requirements may impact the ability of participants to benefit from strategic bidding. Future work should also explore how other non-convex pricing methods currently used by system operators in the United States (see \cite{epri_independent_2019}) may create incentives for adverse self-commitments and self-schedules.

While the induced higher profits by strategic bidding benefit all committed generators, generators who are successful adverse bidders tend to be generators who were already highly profitable under competitive conditions. This could in the long-run bias investment decisions, leading to a resource mix that does not maximize social welfare.

\singlespacing
\printbibliography









\end{document}